\documentclass[12pt,preprint]{aastex}

\shorttitle{Photophysics of SNPs}
\shorttitle{X et al.}

\begin{document}

\title{The Photophysics of the Carrier of Extended Red Emission}
\author{Tracy L.\ Smith\altaffilmark{1} \&  Adolf N.\ Witt\altaffilmark{2}}
\altaffiltext{1}{Department of Physics \& Astronomy, Louisiana State
University, Baton Rouge, LA 70803}
\altaffiltext{2}{Ritter Astrophysical Research Center, The University 
of Toledo, Toledo, OH 43606}

\begin{abstract}
Interstellar dust contains a component which reveals its presence by emitting 
a broad, unstructured band of light in the 540 to 950 nm wavelength range, referred to as Extended Red Emission (ERE). The 
presence of interstellar dust and ultraviolet photons are two necessary conditions for ERE to occur. This is 
the basis for suggestions which attribute ERE to an interstellar dust component capable of photoluminescence. In this study, 
we have
collected all published ERE observations with absolute-calibrated spectra for interstellar environments, where the density of 
ultraviolet photons can be estimated reliably. In each case, we determined the band-integrated  ERE intensity, the wavelength 
of peak emission in the ERE band, and the efficiency with which absorbed ultraviolet photons are contributing to 
the ERE. The data show that radiation is not only driving the ERE, as expected for a photoluminescence process, but is 
modifying the 
ERE carrier as manifested by a systematic increase in the ERE band's peak wavelength and a general decrease
in the photon conversion efficiency with increasing densities of the prevailing exciting radiation. The overall spectral 
characteristics 
of the ERE and the observed high quantum efficiency of the ERE process are currently best matched by the recently proposed 
silicon 
nanoparticle (SNP) model. Using the experimentally established fact that ionization of semiconductor nanoparticles quenches 
their photoluminescence, we proceeded to 
test the SNP model by developing a quantitative model for the excitation and ionization equilibrium of SNPs under interstellar 
conditions for a wide range of radiation field densities. With a single adjustable parameter, the cross section for photoionization, the model reproduces the observations of ERE intensity and  ERE efficiency remarkably well. The assumption that about 50\% of the ERE carriers are neutral under radiation conditions encountered in the diffuse interstellar medium leads to a prediction of the ionization cross section of SNPs of average diameter of 3.5 nm for single-photon 
ionization 
 of $ \leq 3.4 \cdot 10^{-15}$ $\rm cm^{2}$. The shift of the ERE band's peak wavelength
toward larger values with increasing radiation density requires a change of
the size distribution of the actively luminescing ERE carriers through a 
gradual removal of the smaller particles by size-dependent photofragmentation.
We propose that heat-assisted Coulomb decay of metastable, multiply charged
SNPs is such a process, which will remove selectively the smaller components
of an existing SNP size distribution. 

\end{abstract}


\section{Introduction}

Extended Red Emission (ERE) is an interstellar photoluminescence phenomenon, which is widely observed in 
astrophysical environments where 
interstellar dust is exposed to ultraviolet photons (Gordon et al. 1998, and references therein). The emission appears in 
the form of a broad, 
unstructured band, limited in extent to 
the 540 - 950 nm spectral range. The pronounced variability of the peak wavelength of this band 
(610 - 880 nm) and the correlated variability of the FWHM of the 
band (60 - 180 nm) are two defining characteristics of the ERE (Darbon et al. 1999). Recent summaries of the observational 
data concerning the ERE have been 
given by Witt et al. (1998), Smith (2000) and by Ledoux et al. (2001), which should be consulted for the numerous 
observational references.

The peak intensity of the ERE in different environments, ranging from the high-Galactic-latitude 
diffuse interstellar medium (ISM) of the 
Milky Way Galaxy to HII regions, is generally proportional to the density of the illuminating radiation 
field (Gordon et al. 1998). This 
is expected for a photon-driven process involving an interstellar dust mixture, in which the ERE carrier 
represents an approximately constant fraction of the dust mass. 
However, a high degree of spatial variability of the ERE intensity, e.g. the presence of bright ERE 
filaments contrasted by locations without 
any detectable ERE within the same individual objects, is also part of the ERE phenomenon (e.g. Witt \& Malin 1989; 
Witt \& Boroson 1990). While faintest 
on an absolute scale in the diffuse ISM of the Galaxy, the ERE, there, is comparatively intense relative to the density of the 
prevailing radiation field 
(Gordon et al. 1998; Szomoru \& Guhathakurta 1998). The correlation of ERE intensity with HI column 
density at high galactic latitudes has 
led to an estimate of a lower limit to the quantum yield of the ERE process of $10 \pm 3$ \% (Gordon et al. 
1998). This lower limit was
arrived at with the assumption that the ERE carrier is the sole source of absorption of interstellar 
photons in the 90 - 550 nm spectral 
range. Given that almost certainly other dust components contribute to this
absorption, the intrinsic quantum efficiency of the ERE 
carrier must exceed the lower-limit 
estimate in inverse proportion to the fraction of photons actually absorbed by it. 

The realization that the ERE carrier is both a luminescing agent of high intrinsic quantum efficiency ($ \gg$ 10\%) as 
well as a major contributor to the UV/optical absorption by interstellar grains had a profound impact 
on the discussion over the identification of the ERE carrier's nature. If we estimate the intrinsic 
quantum yield to be about 50\%, the ERE carrier intercepts about 20\% of the photons absorbed by interstellar dust in the 90 - 550 
nm range.  In view of this high cross section, the ERE carrier is restricted chemically to 
the small number of relatively abundant but highly depleted elements making up most of the mass of 
interstellar grains, i.e. C, Fe, Si, Mg, 
plus O and H. Even without these new restrictions, many of the previously proposed ERE carrier candidates, which
are laboratory analog materials and models, had encountered serious difficulties. Proposed candidates either lacked the 
ability to match the wide range of ERE spectral variations or they implied
the presence of additional emission features or significant 
continuum luminescence shortward of 540 nm, which is not in agreement with observations.  Witt et al. (1998) 
and Ledoux et al.(2001) present detailed discussions of proposed ERE carrier  candidates and the shortcomings of most.

With the high intrinsic  quantum efficiency added to the observational constraints, the discussion over viable ERE carrier candidates currently has narrowed 
to a choice between
crystalline silicon nanoparticles (Ledoux et al. 1998; Witt et al. 1998; Ledoux et al. 2001) and aromatic 
hydrocarbon clusters (AHC) (Seahra \& Duley 1999). The silicon nanoparticle 
(SNP) model derives its strength from the extensive 
agreement between laboratory-derived experimental data on SNP luminescence properties (e.g. Ledoux et al. 2000) and 
the astronomical ERE data, while relevant
laboratory data for the AHCs currently do not exist for the size range of particles proposed by 
Seahra and Duley (1999). The AHC 
model, thus, is based upon theoretical extrapolations from smaller hydrocarbon molecules. 
Particular weaknesses of the AHC model include the relative constancy of the wavelength of its predicted 
emission peak (Seahra \& Duley 1999, Fig. 2), its prediction 
of secondary features not confirmed by observation (Gordon et al. 2000), and a suggested correlation 
between the strength of absorption 
in the 217 nm UV extinction band and the intensity of the ERE, again at variance with observation.

By contrast, the SNP model can explain the variable peak wavelength of the ERE band  
through the size dependence of 
photoluminescence by oxygen-passivated silicon nanocrystals residing within the quantum-confinement 
regime (Delerue et al. 1993; Ledoux et al. 2000; Takeoka et al. 2000).
Furthermore, the quantum efficiency of individual well-passivated SNPs has been found experimentally to be near 100\% (Credo et 
al. 1999; Ledoux et al. 2001). 
The absorption coefficient per unit volume of nanocrystalline silicon (Theiss 1997; Li \& Draine 2001) is about an order of magnitude 
larger than that of average 
interstellar dust (Huffman 1977), so that the cosmic abundance of Si seems  sufficient to explain the 
observed ERE intensities (Witt et al. 1998) and the required associated UV/optical absorption cross section. Ledoux et al. (2001) have estimated
that the mass of interstellar SNPs would be about 2\% of the mass of all interstellar dust, if the intrinsic quantum efficiency of interstellar SNPs is 50\%. This
would require about 20\% of the interstellar silicon in the case that
the relative abundance of silicon is given by solar abundances, but close to 40\%, if B-star abundances prevail (Snow \& Witt 1996). Furthermore, recent observations
(Gordon et al. 2000) have revealed tentative evidence for the existence of a second, related luminescence band
at 1.15 $\mu$m in a bright ERE filament in the reflection nebula NGC 7023, matching in wavelength and width a similar band seen in SNPs at low temperatures in the laboratory (Meyer et al. 1993; Takeoka 
et al. 2000). An increasingly compelling case can thus be made that SNPs may well be the carriers of the ERE, because 
their experimentally established properties meet the constraints imposed by ERE observations and cosmic abundances. However, new constraints based upon the expected infrared emission characteristics of SNPs (Li \& Draine 2001) have
been proposed recently, which, when combined with suitably sensitive observations, may soon place severe limits on the possible interstellar abundance of oxygen-passivated SNPs or of pure-silicon SNPs.

Both the AHC and the SNP models, just as earlier ERE models based on hydrogenated amorphous carbon solids, such as HAC (Duley 1985) or QCC (Sakata
et al. 1992), rely on across-bandgap recombination of electrons and holes in
a semiconductor structure for the source of emission in the ERE band. The width
of the band, in all instances, is attributed to a distribution of sizes of the luminescent entities with size-dependent bandgaps, while the peak wavelength of the band is related to
the bandgap of the dominant particle size. In this paper we plan to examine whether the photophysics
of such semiconductor nanoparticles under astrophysical conditions can account
for the characteristics of the ERE seen in a wide range of astrophysical environments. We will be guided by experimental results found for SNPs and other semiconductor nanoparticles, and we will use experimentally determined
parameters for SNPs, where available, as input into numerical calculations.

First, however, we begin by characterizing the environments, where ERE has been observed. We compute the UV-radiation densities for these environments and compile data for the wavelengths of peak ERE intensities and band-integrated ERE intensities. We proceed by estimating ERE quantum yields
for locations other than the diffuse ISM. Then, starting from a strong base of
experimentally determined properties of SNPs, we develop a model for the photophysics of the ERE carrier in 
interstellar environments. This model must explain,
in a self-consistent manner, the observed variations of the ERE intensity, the ERE quantum efficiency, and 
the ERE peak wavelength 
as a function of the density and hardness of the local radiation fields. This model will identify the 
photoionization of the ERE carrier as the physical process which controls the fraction 
of carrier particles capable of luminescing in a given
environment, which itself is characterized by the density and spectrum of the prevailing photons and by the density and 
temperature of the free electrons. Finally, this model must lend itself to making specific predictions both with respect to 
future ERE observations as well as
to yet not-measured carrier properties.

In Section 2 of this paper, we accumulate the available observational data, which provide information on ERE 
intensities, ERE quantum 
efficiencies, and ERE-band peak wavelengths in different sources. Critical elements will be the 
estimates of the UV radiation densities 
and the lower limits to the ERE quantum efficiencies, which will be determined from information about 
the spectral
types and luminosities of the illuminating stars, the projected distances of the observed regions from 
the sources in question, and the observed ratios of ERE intensity to scattered light intensity over the same wavelength band.
In Section 3 we introduce evidence showing photoionization to be a possible process controlling the ERE; we 
discuss the ionization mechanisms and their
respective rates, and we present the relevant rate equations and their solutions. Section 4 contains the 
discussion where observational data and model predictions will be compared. This is followed by a summary in Section 5.

\section{The Intensity, Peak Wavelength, and Efficiency of ERE}

\subsection{Data and Observations}

We restricted our data sample to objects and environments where the sources of 
illumination are well-known with respect to spectral type, distance, and luminosity, thus permitting a reliable estimate of the
density of the exciting radiation field. Furthermore, we focused on 
objects where the dust compositions are derived from average interstellar dust rather than being the result of local dust formation. Thus, 
we excluded
existing ERE observations of planetary nebulae (Furton \& Witt 1990, 1992)
and limited ourselves to reflection nebulae, HII regions, dark nebulae,
and the diffuse, high-$\mid$b$\mid$ Galactic cirrus, all in the Milky Way Galaxy.

Twenty-one of the twenty-two reflection nebulae (RN) included here were observed by 
Witt \& Boroson (1990) during a spectroscopic survey of RN, wherein the ERE 
band-integrated intensity and the peak wavelength 
of the ERE band were measured.  For the Orion Nebula, these quantities were
taken from the observations of Perrin \& Sivan (1992).  The 
data for the high-Galactic-latitude dark nebula Lynds 1780 (L 1780) were taken from the spectrophotometric observations of Mattila (1979).  One 
observation of an ERE 
peak wavelength for the Red Rectangle was obtained from Rouan et al. (1995) and one
peak ERE wavelength point for the Bubble Nebula (NGC 7635) was taken from Sivan \& Perrin (1993).  The 
ERE peak wavelength measurement for a high-Galactic-latitude cirrus came from Szomoru \& 
Guhathakurta (1998) and the efficiency and band-integrated intensity of ERE in the ISM were
taken from Gordon et al. (1998).

\subsection{Dependence of ERE Properties on Radiation Field}

Gordon et al. (1998) demonstrated that in a multitude of ERE sources the ERE band-integrated intensities and the intensities of the scattered light in the same 
wavelength band are roughly proportional to each other over several orders of magnitude in intensity. 
The scattered light intensity is determined
by the column density of scattering grains and the density of the illuminating radiation in the wavelength range of the ERE band. This finding, therefore, suggests that, on 
average, the ERE carriers
represent a fixed fraction of the dust present and that the closely related illuminating radiation field at wavelengths shorter than the ERE band is the source of ERE excitation. To quantify this 
relationship, an estimate of the densities of the exciting radiation fields is essential.

An analysis of ERE intensity variations in reflection nebulae in relation to the degree of internal reddening as a function of position within these nebulae by Witt \& Schild (1985) and the non-detection of ERE in reflection nebulae illuminated by stars cooler than 7000 K (Darbon et al. 1999; Witt \& Rogers 1991) suggest that the excitation of the ERE carrier is dominated by photons
in the wavelength range shortward of 250 nm. We estimated the relevant radiation field density in the diffuse ISM, $U_{isrf}$, by
integrating over the  flux values of Mathis et al. (1983) from 91 to 250 nm at a distance of 10 kpc from the Galactic 
Center. All other radiation field 
densities will be normalized to this reference value, $U_{isrf}$ = 9.7$\cdot 10^{-14}$ ergs $\rm cm^{-3}$.

Then, for nebulae with known illuminating stars, we
calculated the radiation field density at the top of the illuminating star's atmosphere,
$U_{star}$, using the star's temperature and the appropriate atmosphere
model (Kurucz et al. 1974) for the wavelength range from 91 to 250 nm.  
This radiation field density was then corrected for geometric dilution, 

\begin{eqnarray}
U_{obs} & = & \frac{U_{star}}{2}(\frac{r}{D})^{2},
\end{eqnarray}

\noindent where $U_{obs}$ is the radiation field density at the point of ERE observation, 
\emph{r} is the stellar radius and \emph{D} is the projected distance of the ERE observation from the
star in the plane of the sky.  The radiation field strength for each point of 
observation was then expressed in units of $U_{isrf}$,

\begin{eqnarray}
U & = & \frac{U_{obs}}{U_{isrf}}.
\end{eqnarray}

Since the values of \emph{D} are measured in the plane of the sky, they are
lower limits to the true distances of the observed positions 
from the stars, and \emph{U} is an upper limit to the actual radiation field strength.  
Figure 1 shows a plot of the accumulated ERE intensity values versus the associated 
densities of the radiation field within specific and among different RN, the Orion Nebula, L 1780, the diffuse ISM, and the Red Rectangle. The 
diffuse ISM point is an average of many lines of sight (Gordon et al. 1998).
The downward-pointing error bars on some of the RN points 
indicate upper limit determinations for the ERE intensity by Witt \& Boroson (1990). The Red Rectangle is included here to indicate its 
exceptional ERE brightness, but we note that the dust in this bi-polar proto-planetary nebula is most likely
of local origin and thus is not representative of the typical interstellar dust mixture. Figure 1 confirms that the different ERE 
intensities in the diffuse ISM, in RN and in the Orion Nebula are a result of scaling the density of the exciting radiation
fields, as expected for a photoluminescence process. Specifically, we expect
the ERE intensity to exhibit a proportionality given by

\begin{eqnarray}
I(ERE) & \propto & U \cdot \eta \cdot (1 - exp(-\tau_{ERE})),
\end{eqnarray}

\noindent where $\eta$ is the ERE photon conversion efficiency, defined in the following section, and where $\tau_{ERE}$ is the effective optical depth for dust absorption
at the wavelength of the ERE. The examination of Figure 1 reveals that the
detected ERE intensities range over four orders of magnitude, while the
corresponding exciting radiation field densities vary by about six orders of magnitude. Given that the values of $\tau_{ERE}$ are comparable in the different
ERE sources, Figure 1 provides the first qualitatative evidence for substantial
variations in the ERE efficiency among different environments. In particular,
the Red Rectangle, illuminated by a relatively cool B 9.5 III star, must derive its exceptional 
ERE intensity, when compared with the
more typical RN, from a combination of high intrinsic efficiency and relative overabundance of ERE carriers, most likely produced in the local mass outflow.

\begin{figure}[th]
\begin{center}
\epsscale{0.7}
\plotone{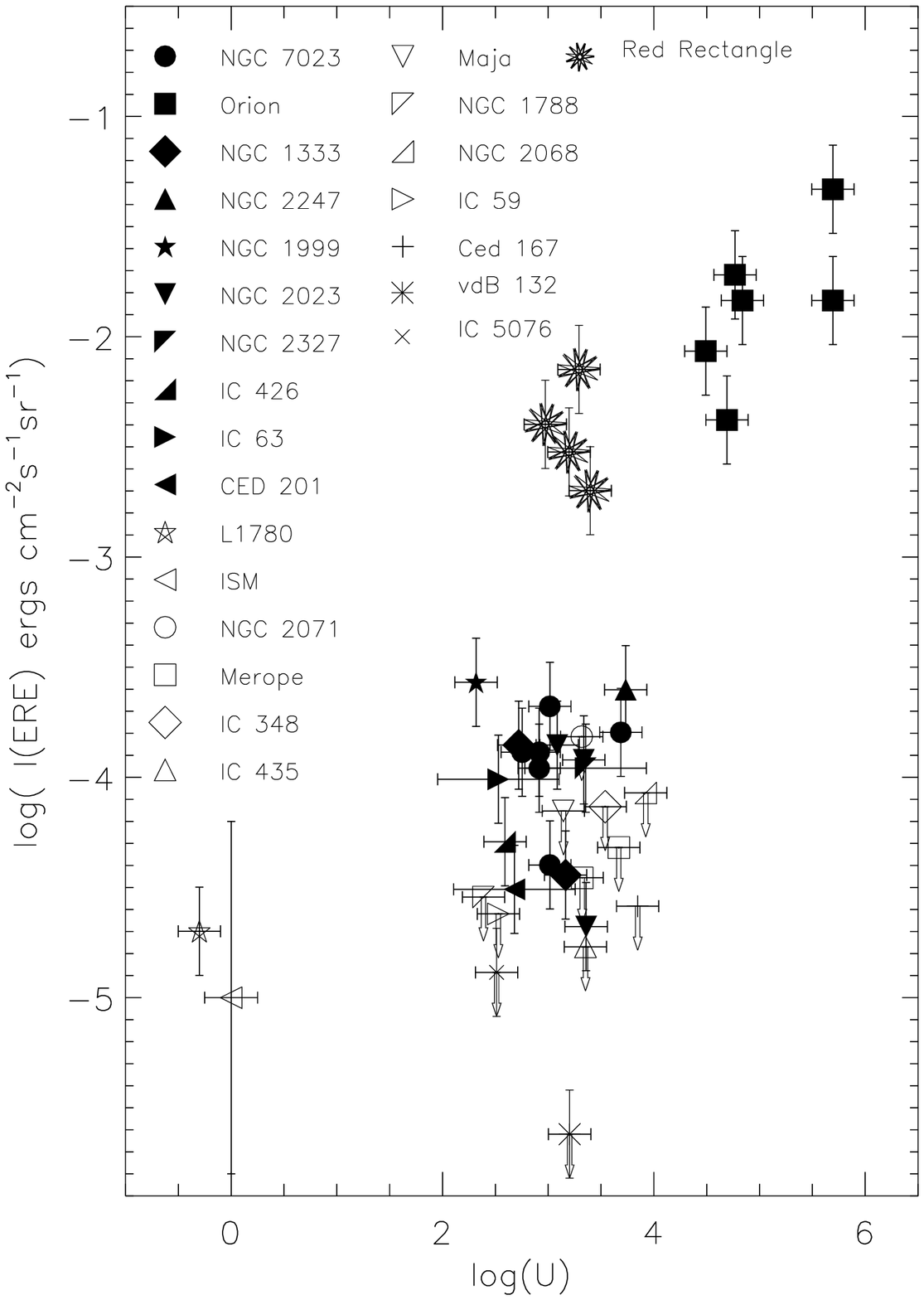} 
\end{center}
\caption{The relationship between the ERE intensity and the density of the radiation 
field in a variety of ERE sources. Downward pointing arrows mark measured upper limits in ERE intensities.}
\end{figure}

Figure 2 shows the relationship between ERE-band peak wavelength, 
$\lambda_{p}$, and radiation 
field density.  Again, due to the projection effect, the 
radiation field strengths calculated for the RN, Orion Nebula, and the Red 
Rectangle are strict upper limits. Figure 2 indicates that the longest peak wavelengths  for the ERE band are 
found in high radiation density environments, and the shortest are found in the lowest radiation density
environments, as represented by the high-Galactic-latitude cirrus. The high-$\mid$b$\mid$ dark nebula L 1780 appears to take on an outlier role. We will 
comment on the likely explanation for this effect in Sect. 4.3.

\begin{figure}[th]
\begin{center}
\epsscale{0.9}
\plotone{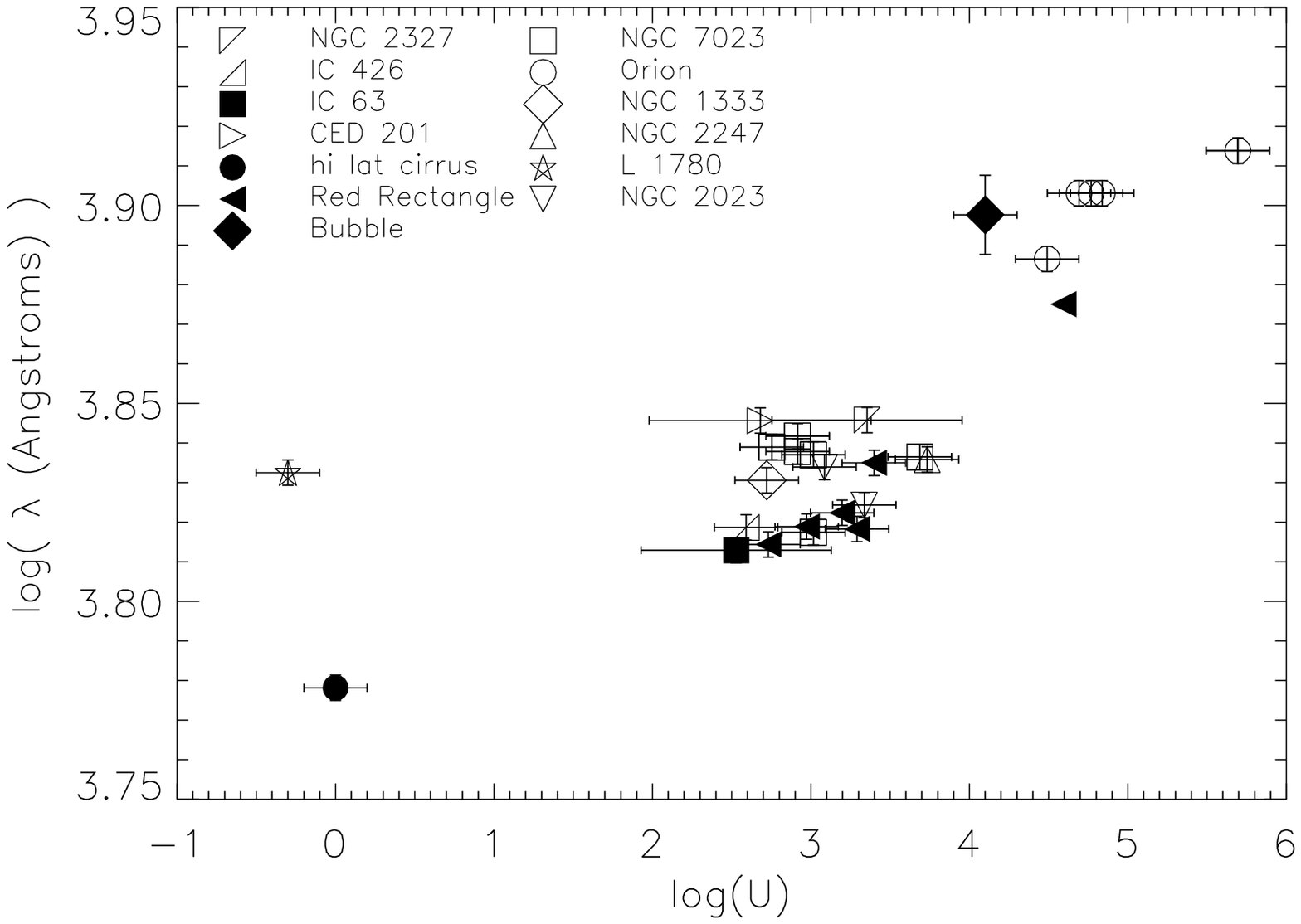}
\end{center}
\caption{The peak wavelength of ERE plotted as a function of radiation field density.}
\end{figure}

\subsection{The Photon Conversion Efficiency of the ERE Process}

One can estimate a lower limit to the ERE conversion efficiency, if both
the number of emitted ERE photons and the number of UV/optical photons 
absorbed within the same environment are known. It is assumed that the
photons absorbed by dust which can possibly lead to the emission of an
ERE photon are in the wavelength range between 91 nm (ionization of H) and 540 nm (edge of ERE
band), although 250 nm is probably a more realistic upper limit of this range
(see Sect. 2.2).  The efficiency determined in this manner is a lower limit only,
because other dust components, which are not contributing to the ERE, 
may participate in the photon absorption and the ERE excitation may occur
primarily in the short-wavelength portion of the assumed wavelength window of
exciting radiation.

Spectroscopic observations of reflection nebulae show the ERE band 
superimposed on a scattered light continuum.  The scattered intensity
beneath the ERE band, $I_{sca}$(band), is a direct probe of the number of 
illuminating photons within the ERE band and the number of dust grains present in the observed volume giving rise to the 
ERE.  Given the knowledge of (1) the illuminating star's spectral energy distribution, which relates the number of illuminating and scattered photons in the ERE band to the number of photons capable of exciting the ERE, 
(2) the wavelength dependence of extinction and (3) the wavelength dependence of
the dust albedo, the number of photons absorbed between 91 nm and 540 nm,
$N_{abs}$(91-540 nm), can be estimated.  The photoluminescence efficiency 
can then be found as

\begin{eqnarray}
\eta & = & \left(\frac{N_{ERE}}{N_{sca}(\rm band)}\right)\left(\frac{N_{sca}(\rm band)}
{N_{abs}(91-540 \rm nm)}\right)
\end{eqnarray}

\noindent where $N_{ERE}/N_{sca}$(band) is the ratio of the number of ERE photons and
scattered photons, each integrated over all solid angles. This ratio is not
observationally accessible to a spatially fixed observer. Instead, we measure
the intensity ratio $I_{ERE}/I_{sca}$(band), which depends strongly on the range of the dominant scattering angles in a given nebula, defined by the unknown nebular geometry and the direction to the observer. The direction dependence of
the intensity ratio is caused by the fact that the scattered light intensity
is subject to a highly asymmetric phase function, while the ERE is expected to be emitted isotropically. This expectation is supported by experimental evidence, showing colloidal SNPs to exhibit no polarization memory (Koch et al. 1996). Multiple phonon-exciton interactions during the long radiative lifetime
of SNPs as well as the rapid rotation of free-flying SNPs effectively destroy
all directional relations between exciting photons and emitted ERE photons.
This assessment is confirmed by the observed ERE band intensities and the corresponding scattered light intensities seen in RN
(Witt \& Boroson 1990, Fig. 1). For a given scattered light intensity, the ERE intensities in RN vary by about one order of magnitude at most, while for a given ERE intensity the corresponding scattered light intensities vary by nearly two orders of magnitudes. We hope to overcome this basic problem by averaging over multiple observations in a given RN and by
averaging the data for a relatively large sample of RN representing a range of
geometries, and with this reservation set $N_{ERE}/N_{sca}$(band) = $I_{ERE}/I_{sca}$(band).

The ratio of photons scattered within the wavelength range of the ERE band to
photons absorbed within the wavelength range 91 - 540 nm, $N_{sca}$(band)/$N_{abs}$(91-540 nm), was determined with the help of a Monte Carlo radiative transfer code. This code included the local dust opacity law, characterized by the value of the ratio of total to selective extinction, $R_{V}$, the wavelength dependence of the dust albedo and phase function, and the spectral energy distribution of the exciting star, characterized by its
effective temperature. The principal source of uncertainty, again, is the unknown geometry of the RN, which can be different in each instance. For simplicity, we adopted a spherical geometry with a centrally embedded star.
The optical depth of the radius (1 pc) was taken to be 0.5 at the ERE 
wavelength and the density of the nebulae was assumed uniform. The absolute surface brightness and radial surface brightness distribution produced with 
such models agree well with measured surface brightness data for bright
reflection nebulae (Witt \& Schild 1986). In the model output, we examined the ratio $N_{sca}(\rm band)/N_{abs}(91-540$ $\rm nm)$ at an intermediate angular offset corresponding to 0.3 of the projected radius of the nebula. This arrangement covers a large range of scattering angles but still favors forward-scattering, appropriate for a collection of objects which include many of the RN with the highest-known surface brightnesses. Calculations were performed for two values of the UV albedo, 0.4 and 0.6, to correspond to limiting cases of low $R_{V}$/strong 218 nm absortion feature and high 
$R_{V}$/weak 218 nm absorption feature, respectively. The dust albedo in the ERE band was set at 0.6, consistent with numerous observational determinations (Witt \& Gordon 2000). The phase function asymmetry was assumed to increase with
decreasing wavelength, e.g., g = 0.6 at the ERE wavelength vs. g = 0.7 at
the wavelength of the exciting UV radiation. An example of the model results 
for the ratio

\begin{eqnarray}
\frac{I(ERE)}{I_{sca}(\rm band)} & = & \eta \cdot \frac{N_{abs}(91-540  \rm nm)}{N_{sca}(\rm band)}
\end{eqnarray}

\noindent is shown for $\eta$=0.01 as a contour plot in Figure 3, for 
$a_{UV}$=0.6, 2.5$\le R_{V} \le 5.0$, 
and $1\cdot 10^{4} \rm K \le T_{eff} \le 3.5\cdot 10^{4} \rm K$, 
corresponding to stars of approximate spectral types A 0 to O 7.  Figure 3 illustrates the fact 
that $I_{ERE}$/$I_{sca}$ can change easily over a factor of 30 to 40 for the same value of the 
quantum efficiency $\eta$, depending upon the values of $R_{V}$ and $T_{eff}$. 
Correct estimates for the nebular opacity law and the spectral type of the illuminating stars are, therefore, of critical importance. Table 1 shows the values we adopted for the nebulae with ERE detections and well-determined upper limits. 

\begin{table}
\begin{center}
\caption{Data Related to Efficiency Estimates \label{tbl-1}}
\begin{tabular}{lllll}
Object & Star Sp. T.& $R_{V}$ & $I_{ERE}$/$I_{sca}$(band) & 
$\eta$(\%) \\ \hline \hline
NGC 1333     & B8 V	       & 4.7	     & 0.07 - 0.16	& 1.8 - 4.0   \\
NGC 1788 & B9 V & 4 & $<$ 0.03 & $<$ 1.04 \\
NGC 1999 & A0 & $\sim$ 3.6 & 0.06	& 2.0 \\
NGC 2068 & B2 II - III & 4 & $<$ 0.03 & $<$  0.16 \\
NGC 2023       & B1.5 V        & 4.11       & 0.02 - 0.42	& 0.04 - 1.2         \\
NGC 2071 & B5 V & 4.5 & $<$ 0.1 & $<$ 1.2 \\
NGC 2247             & B3 Pe        & 3.25       & 0.08	& 0.4    \\
NGC 2327  & B1        & 3.1 - 4.5       & 0.06	& 0.1  \\
NGC 7023   & B3 Ve        & 3.2       & 0.04 - 0.15	& 0.2 - 0.6  \\
IC 59 & B0 V & 3.5 & $<$ 0.16 & $<$ 0.19 \\
IC 63    & B0 IVe        & $\sim$ 3.5	     & 0.68	& 4.7   \\
IC 348 & B5 V & 3.5 & $<$ 0.05 & $<$ 0.4 \\
IC 426        & B8 p        & $\sim$ 4       & 0.21	& 7.9   \\
IC 435 & B5 V & 5.3 & $<$ 0.01 & $<$ 0.13 \\
IC 5076 & B8 Ia & 3.2 & $<$ 0.09 & $<$ 2.48 \\ 
CED 167 & B6 V & 3.2 & $<$ 0.01 & $<$ 0.13 \\
CED 201      & B9.5 V  & 3.42       & 0.02	& 0.5  \\
Maja & B7 III & 3.2 & $<$ 0.02 & $<$ 0.32 \\
Merope & B6 IV & 3.6 & $<$ 0.01 & $<$ 0.01 \\
VdB 132 & B3 V & 4 & $<$ 0.02 & $<$ 0.08 \\
Orion Neb.	& 06, 07, 2xB0 & 5.5	& 0.20 - 0.64 	& 0.4 - 1.3 \\ 
L1780	& ISRF	& 3.1	& 0.6	& 13  \\
ISM & ISRF & 3.1 & 0.2 - 2 & 10 \\ \hline \hline
\end{tabular}
\end{center}
\end{table}

The uncertainties introduced by 
the range in UV dust albedos depend on the temperature of the illuminating 
stars
and range from 15\% to 25\% for stars ranging from late B to late O spectral
type. The uncertainties go in the direction of increasing the efficiency
for a given ratio $I_{ERE}/I_{sca}(band)$ for a higher UV albedo. While the overall uncertainties of the derived efficiencies for individual RN 
observations are difficult to assess for reasons discussed above, 
we estimate the uncertainty
of the efficiency of the $\emph{average}$ of all RN detections to be about a factor of two.

\begin{figure}[th]
\begin{center}
\epsscale{0.9}
\plotone{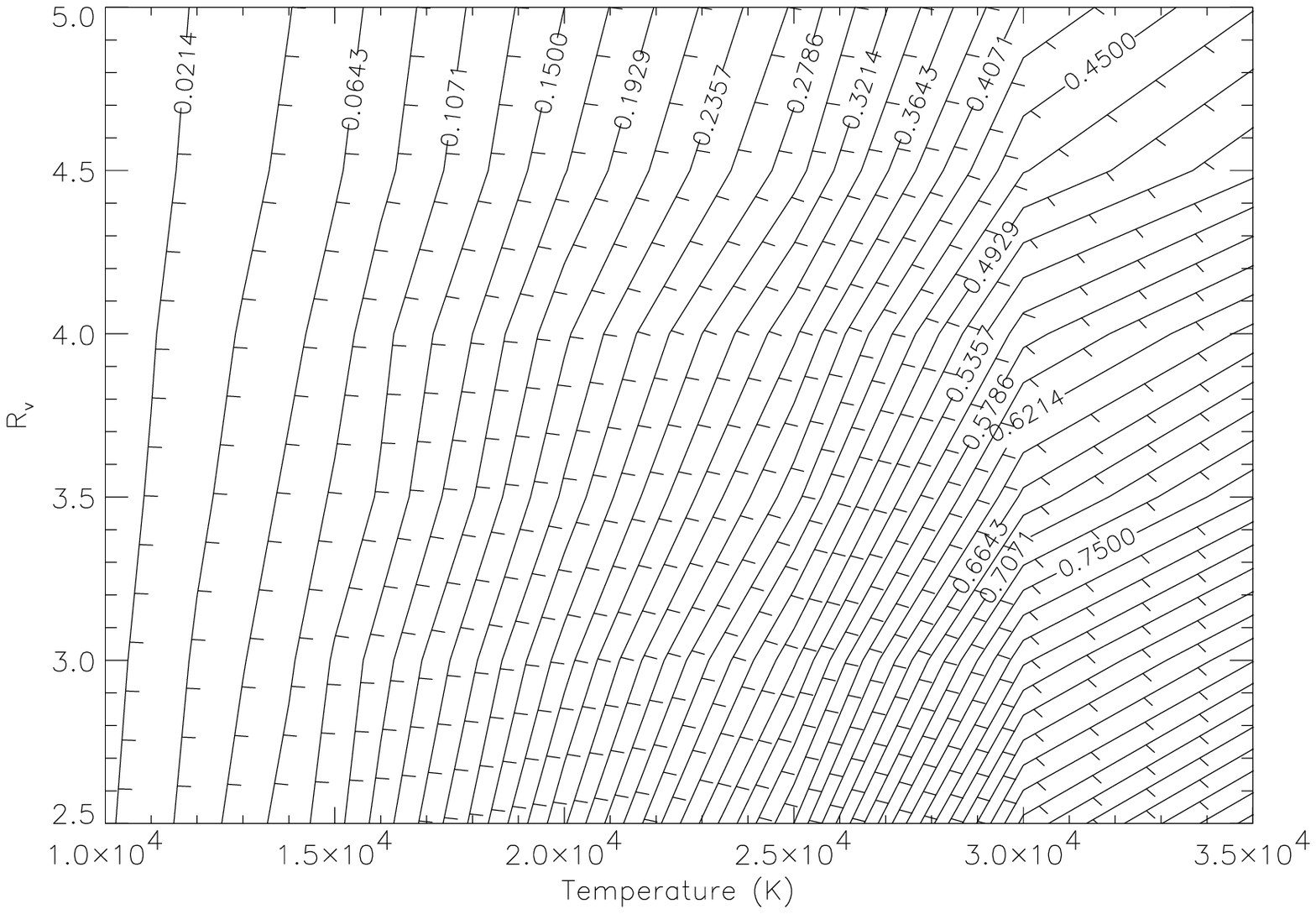}
\end{center}
\caption{A contour diagram showing $I_{ERE}/I_{sca}(band)$ as a function of star temperature and $R_{V}$ for a UV albedo $a_{UV}$ = 0.6, 
and for an assumed ERE efficiency of 1\%.}
\end{figure}

The estimated efficiencies, as defined above, are plotted in Figure 4 as a function of the UV radiation density. The highest efficiencies found are associated with the lowest radiation densities. Also, among reflection nebulae, illuminated by stars between spectral types A 0 and B 0, we note a significant decline of the ERE efficiency by about two orders of magnitude as the radiation density increases by a similar factor. The Orion Nebula data do not follow this 
general trend, a fact that needs to be explained by our model (\emph{c.f.} Sect. 3.4). We do not include
the Red Rectangle data in this graph, because the dust mixture in this object
is not derived from interstellar dust but is locally produced. We will discuss the efficiency of the ERE carriers in the Red Rectangle below.

\begin{figure}[th]
\begin{center}
\epsscale{0.9}
\plotone{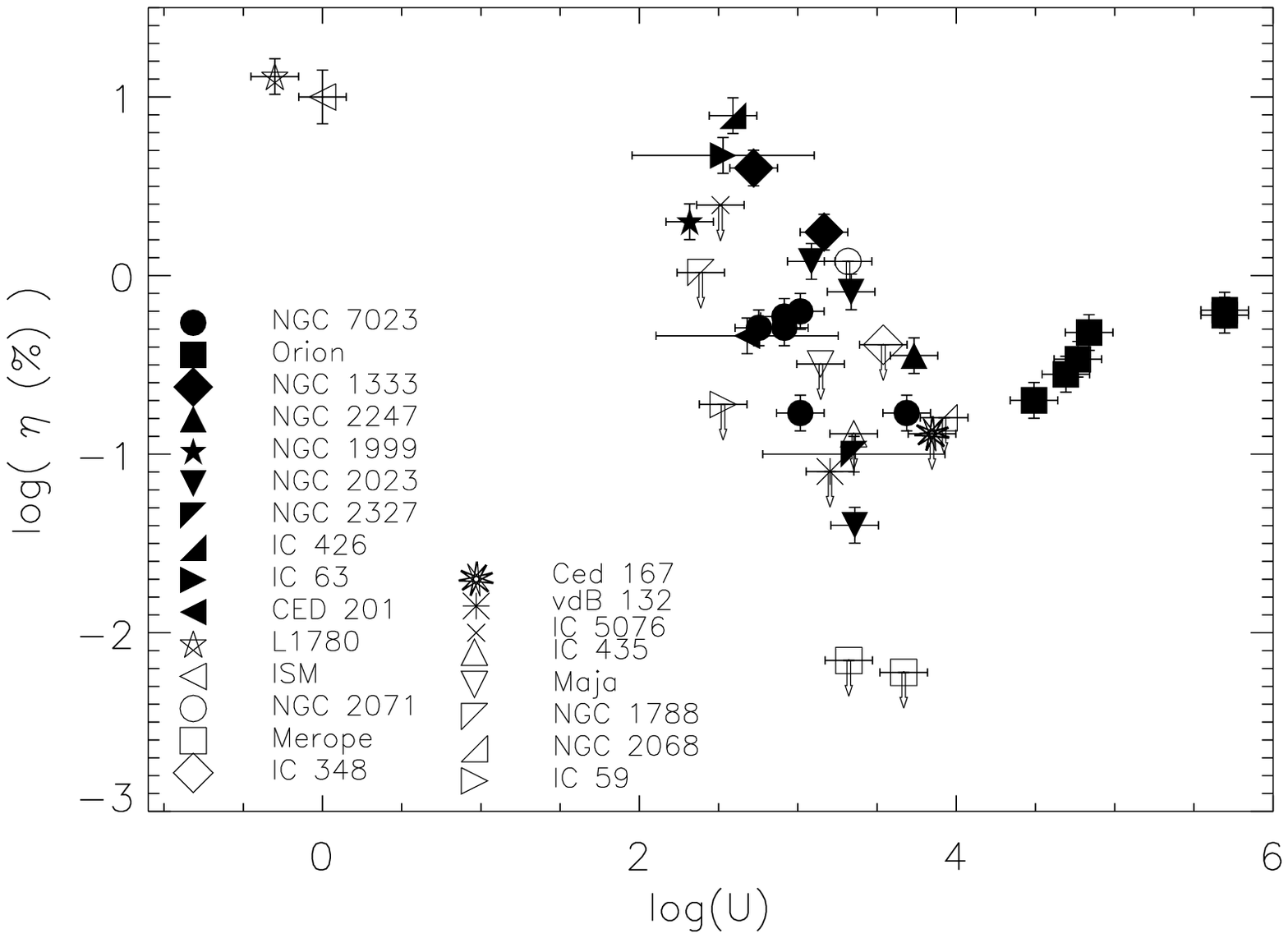}
\end{center}
\caption{The ERE efficiency for a variety of sources plotted as a function of
the radiation field density. Both detections and upper limits are included.}
\end{figure}

\subsection{The Red Rectangle}

The Red Rectangle occupies a special role in the problem of understanding the ERE. Given the relatively cool illuminating star in the object, the exceptionally high ERE intensity suggests that the ERE carriers are present here with an unusually high relative abundance, most likely as a result of local production. By employing the same method as used for RN, we estimate an efficiency for the photon conversion of $\sim$ 20\%, the highest value seen in any object.
The principal source of uncertainty here is the unusual opacity law characterizing the matter in the Red Rectangle (Sitko et al. 1981), which is unlike that of any interstellar or nebular dust. There is only a very weak indication of absorption at the interstellar extinction band centered at 217.5 nm, but
massive absorption shortward of 180 nm is present instead. The ERE is strongest along the
X-shaped walls of the bi-polar outflow cone (Schmidt \& Witt 1991), while the
unidentified infrared band (UIB) at 
3.3 $\mu$m is found to be weak to non-existent along the walls or interface
areas, but very strong in the dust ring surrounding the central binary star (Kerr et al. 1999). Additionally, all of the remaining UIBs 
at 6.2, 7.7, 8.6 and 11.3 $\mu$m have been observed in the Red Rectangle 
(Russell et al. 1978).  Waters et al. (1998) have used observations from the 
Infrared Space Observatory (ISO) to confirm the existence of crystalline 
silicates in a circum-binary disk in the nebula, which suggests a mixed chemistry
for this object.

\section{The Photoionization of Semiconductor Nanoparticles Under Astrophysical Conditions}

\subsection{The Silicon Nanoparticle Model}

Before discussing the photoionization of semiconductor nanoparticles in general, it is essential to briefly summarize the main 
characteristics of the SNP model as representative of a possible carrier of the ERE. 
Silicon is an indirect-bandgap semiconductor material with a bandgap energy of 1.17 eV at 0 K. The 
indirect-bandgap nature implies that an excited electron located in the conduction band needs to undergo 
a change in momentum state before it can recombine with a hole in the valence band. This process, which involves 
electron-phonon interactions, has a 
long time constant ($\sim 10^{-5}$  -  $10^{-3}$ s; Takeoka et al. 2000). Consequently, in the absence of spatial confinement  and with an abundance of non-radiative 
recombination routes available in bulk silicon, photoluminescence in bulk crystalline silicon is a highly unlikely process. Several 
important changes occur when a transition to nanocrystals of silicon is considered (Freedhoff \& 
Marchetti 1997). First, size-dependent quantum confinement of the electron-hole exciton causes the bandgap to widen, shifting it from 
1.17 eV to values consistent with the wavelength range of the ERE band (Ledoux et al. 2000). Second, the spatial 
confinement of the electron forces its wavefunction to spread 
in momentum space in accordance with the Heisenberg uncertainty principle, greatly enhancing the 
probability of radiative recombination with the hole in the valence band without   phonon collisions. Thus,
efficient photoluminescence at energies much higher than 1.17 eV becomes possible. However, this occurs only, provided that all dangling Si bonds at the surface of the nanocrystal are passivated, e.g.,  by hydrogen or oxygen atoms.
Observational constraints suggest that interstellar Si nanocrystals would be passivated by oxygen (Witt et al. 1998). For a range of 
diameters from 1 nm to 7 nm, the peak energy of photoluminescence of SNPs varies between 1.9 eV and 1.2 eV 
(Takeoka et al. 2000), in agreement with the wavelength range over which ERE has been observed in 
astronomical sources. And third, the small size of 
the nanocrystal makes it much easier to arrive at a defect-free structure.
The presence of defects would otherwise provide for radiationless recombination
routes, which would be detrimental to a high quantum efficiency. As a consequence, the resulting well-passivated, nanocrystalline system is a 
particle capable of photoluminescence with essentially 100\% 
efficiency,  emitting a luminescence photon upon each excitation by a photon with an energy exceeding that 
of the bandgap of the particle in question (Credo et al. 1998). An appropriate size distribution of such particles can then 
account for the observed interstellar luminescence seen as ERE, provided suitable shorter-wavelength 
photons are present to cause the excitation.

Secondary, but equally important, aspects of the SNP model concern a plausible mode of formation under astrophysical conditions 
and adequate cosmic abundance of silicon to explain the observed ERE intensities. Witt et al. (1998) 
suggested that the formation of interstellar SNPs could occur as a result of the nucleation of SiO molecules 
in oxygen-rich stellar outflows (Castro-Carrizo et al. 2001; Roche et al. 1991;
Hirano et al. 2001; Garray et al. 2000; Zhang et al. 2000; Rietmeijer et al. 1999; Gail \& Sedlmayr 
1999). To lead to SNPs, the nucleation of SiO must be
followed by annealing and phase separation into an elemental silicon phase in the core and a passivating mantle of $\rm SiO_{2}$. In fact, the $\rm SiO_{2}$ phase
is thermodynamically more favorable for solid silicon oxides. Recent
experiments involving the evaporation and recondensation of SiO lend strong support to this suggestion (Rinnert et al. 
1998, 1999; Murakami et al. 1998). The nucleation of SiO molecules is a 
proposed first stage of the formation of silicate grains (Gail \& Sedlmayr 1999), although the 
theoretical understanding of this formation process is still far from complete.
Given the high mass absorption coefficient of nanocrystalline silicon (Theiss 1997) and the high quantum 
efficiency of SNPs, it has been estimated that the mass of interstellar SNPs amounts to only 1\% to 2\% of the interstellar 
dust mass (Ledoux et al. 2001), while silicon contributes about 8\% of the total mass of interstellar 
solids condensed from a gas of solar composition (Whittet 1992, p.52).

The absorption of interstellar photons by SNPs occurs primarily in the mid- and far-ultraviolet (Zubko et al. 1999; Theiss 1997), which is consistent with astronomical assessments of the excitation of the ERE 
(Witt \& Schild 1985; Darbon et al. 1999). The expected oxide coating of SNPs not only is essential for 
effective passivation, assuring a high overall photoluminescence efficiency, it also contributes to 
the interstellar Si-O stretch absorption band at 9.7 $\mu$m and to a possibly prominent emission feature at 20 $\mu$m (Li \& Draine 2001). Also, the oxide
coating assures that even the smallest SNPs do not luminesce shortward of
540 nm by creating surface-related electronic states within the growing band
gap of such particles (Wolkin et al. 1999). Finally, given that only about 25\% of the energy of an absorbed typical ultraviolet photon is 
used to power the ERE, the bulk of the absorbed energy goes towards heating the SNPs (Duley 1992; Li \& Draine 2001). 
As a consequence of the small size of the SNPs, temperature fluctuations may raise the temperatures of the particles 
temporarily into the range from 50 K to 300 K (Li \& Draine 2001), causing them to contribute to  
thermally driven band and continuum dust emission, mainly in the 15 - 60 $\mu$m wavelength region.

\subsection{ Photoionization Control of ERE Carrier Photoluminescence}

The observational data summarized in Sect. 2 strongly suggest that the 
ultraviolet photon field present in different ERE sources not only powers the ERE (Figure 1), but also 
controls the size distribution of luminescing particles (Figure 2) and the fraction of absorbing 
particles capable of luminescing (Figure 4). Two mechanisms must be 
considered: (i) partial, size-dependent photodestruction and (ii), size-dependent quenching of the photoluminescence by photoionization. Typical SNPs with spectra matching those of the 
ERE have between 200 and 6000 silicon atoms and their instantaneous
photodestruction by thermal heating is a very unlikely process. 
The maximum temperatures of SNPs in typical ERE sources predicted by Li \& Draine (2001) are in the 300 K - 500 K range for SNPs  with diameters of 2nm, which is too low for thermal evaporation. An alternative process is fragmentation by Coulomb explosions of multiply charged
SNPs. Such fragmentation, mostly by ejection of singly-charged monomers and dimers, has been seen experimentally in multiply-ionized, free silicon nanoparticles by Ehbrecht \& Huisken (1999) and Bescos et al. (2000). It is not known, how the presence of an oxide shell affects this disintegration process.
However, there is considerable experimental evidence that single ionization of semiconductor nanocrystals, including SNPs, effectively quenches their photoluminescence (Nirmal \& Brus 1999; Banin 
et al. 1999; Efros \& Rosen 1997; Nirmal et al. 1996; Kharchenko \& Rosen 1996; Wang et al. 2001). If an ionized nanocrystal is photoexcited, 
the photogenerated electron-hole pair recombines by transferring its energy to the strongly Coulomb-coupled second hole via a non-radiative Auger decay process (Nirmal \& Brus 1999). Partial ionization of SNPs is expected under conditions existing in the diffuse ISM and a higher degree of ionization, including multiple ionization, is expected in regions of higher radiation density
(Bakes \& Tielens 1994; Weingartner \& Draine 2001). The threshold for single-photon ionization of free silicon clusters with about 200 silicon atoms has been found experimentally to lie near 5.1 eV 
(Fuke et al. 1993). We expect the ionization to be more difficult, if not requiring a higher energy, when a passivating oxide shell is present, but no experimental data appear to exist for this case. The energy per exciting photon of most laboratory photoluminescence experiments is 
less than the ionization threshold value. Consequently, the ionization process usually seen under laboratory 
conditions is Auger ionization (Chepic et al. 1990; Kovalev et al. 2000b), which arises when a second 
electron is excited into the conduction band before the first electron has completed its recombination 
with its hole and when the combined
excitation energies of the two electrons exceed the ionization threshold.
The strong Coulomb interaction between the two excited electrons causes one to recombine with a hole while the other 
leaves the system.
The Auger process is both highly efficient in quantum-confined nanocrystals, and  particularly 
important in SNPs as a result of the long photoluminescence lifetime of singly-excited electrons residing in 
the conduction band, which ranges up to msec in duration, but is dependent on size and temperature of the SNPs (Takeoka et al. 2000). 

Under interstellar conditions, the photon spectrum extends to 13.6 eV, and photoionization by single 
photons must clearly be the dominant process in the diffuse ISM, given the high degree of dilution of the interstellar 
radiation field. As we will show below, despite the fact that the rate of the two-photon Auger ionization 
process increases with the square of the radiation field density, the two ionization processes are not likely to reach comparable rates in any of the radiation environments, where ERE is observed.

As shown in Figure 4, the observed lower limit of the ERE quantum yield in the diffuse ISM is about $10 \pm 3$ \% 
(Gordon et al. 
1998). After making allowance for other grain components contributing to the absorption of UV/optical 
photons without contributing to the ERE, we may estimate that the intrinsic quantum efficiency of the 
ERE carrier particles could be near 50\% (Witt et al. 1998). If photoionization controls the ability 
of the ERE carrier to luminesce, the intrinsic quantum efficiency then is
proportional to the non-ionized fraction of the carrier
particles. Thus, we find that the degree of ionization of the ERE carrier particles in the diffuse ISM must be fairly low, not exceeding 50\%. The degree of ionization could be less, if a 
certain fraction of the carriers is otherwise defected and thus unable to luminesce, because the structure of 
such defected nanoparticles permits radiation-less recombination of electrons and holes.
These expectations regarding the ionization state of nm-sized ERE carrier particles are in good agreement
with the predictions of likely charge states
for particles of 3 nm diameter in the cold, diffuse ISM by Bakes \& Tielens (1994) and Weingartner \& Draine (2001)
 
The expected low degree of ionization in interstellar space places a strict
upper limit on the ionization cross section of SNPs for single-photon ionization, assuming that SNPs cause the ERE. We will 
calculate this cross section below. Since no measurements exist for this cross section, our calculation will 
represent a specific prediction made by the SNP model, which later laboratory experiments may test.

\subsection{The Photoionization Equilibrium Rate Equations and their Solutions}

Let us assume that interstellar SNPs are perfectly passivated and that the
only way to quench their photoluminescence is by photoionization. We further assume the absorption
characteristics of SNPs to be unchanged, whether or not the latter are ionized. Then, under interstellar 
conditions, the photoluminescence properties of SNPs are controlled by the probabilities of occupation 
of four distinct states.
The first is $P_{0}$, the probability that a SNP is in the neutral ground state, with all electrons in 
the valence band. It is seen easily that $P_{0}$ measures the intrinsic photoluminescence quantum 
efficiency of the total SNP ensemble, consisting of neutral and ionized particles.
The second essential probability is $P_{1}$, which measures the fraction of SNPs in the neutral, excited 
state in which one electron is momentarily in the conduction band. Since the decay from this state 
almost invariably is via the emission of a photoluminescence photon, $P_{1}$ is directly proportional to 
the ERE intensity produced by the SNPs. A third state, measured by the probability 
$P_{2}$, represents neutral SNPs 
in which a second photoexcitation has occurred before the first excitation had enough time to decay. 
The most likely result of the strong Coulomb interaction of the two electrons is Auger ionization on 
a timescale of typically
$\tau_{A} \sim 10^{-10}$ s (Delerue et al. 1998), rendering the SNP unable to luminesce. However, under astrophysical conditions this probabilty is expected to be very small. The fourth state of importance is the ionized 
state, measured by the probability $P_{+}$. SNPs can enter this state via either one of the two ionization processes. We will 
consider only the once-ionized state here, although multiply-ionized SNPs are likely in 
environments with high photon densities and comparatively low densities of free electrons. We ignore the possibility of negatively charged SNPs, because the
rate of photoionization of negatively charged SNPs is expected to be large
compared to the electron capture rate of a neutral SNP, given the radiation and electron densities in ERE-emitting environments. Our model accounts for the presence of higher positive charge states by increasing the value of $\tau$, the lifetime of the ionized state. We will 
discuss the role of multiple ionization in SNPs in Section 3.5 of this paper.

We can formally express the time rate of change of the four probabilities
$P_{0},  P_{1},  P_{2}$, and  $P_{+}$, by the following rate equations (Chepic et al. 1990):
\begin{eqnarray}
\dot{P_{0}} = -W_{1} \cdot P_{0} - \frac{P_{0}}{\tau_{i}} + \frac{P_{1}}{\tau_{1}}
\end{eqnarray}
\begin{eqnarray}
\dot{P_{1}} = W_{1} \cdot P_{0} + \frac{P_{2}}{\tau_{2}} - \frac{P_{1}}{\tau_{1}} - W_{2} \cdot P_{1} + 
\frac{P_{+}}{\tau}
\end{eqnarray}
\begin{eqnarray}
\dot{P_{2}} = W_{2} \cdot P_{1} - \frac{P_{2}}{\tau_{2}} - \frac{P_{2}}{\tau_{A}}
\end{eqnarray}
\begin{eqnarray}
\dot{P_{+}} = \frac{P_{2}}{\tau_{A}} + \frac{P_{0}}{\tau_{i}} - \frac{P_{+}}{\tau}
\end{eqnarray}
and, consistent with our assumptions, the following normalization:
\begin{eqnarray}
P_{0} + P_{1} + P_{2} + P_{+} = 1.0
\end{eqnarray}
To solve these equations for any given environment, we assume steady state:           
\begin{eqnarray}
\dot{P_{0}} = \dot{P_{1}} = \dot{P_{2}} = \dot{P_{+}} = 0
\end{eqnarray}

The factors $W_{1}$ and $W_{2}$, respectively, represent the rates of photoexcitation 
of SNPs in the neutral ground state and the neutral, once-excited state
by the prevailing radiation field, such that the combined energy of the two
photons causing the excitations exceeds the ionization potential of the SNPs.
For our subsequent calculations we will assume $W_{1} = W_{2}$, which is approximately correct (Efros \& 
Rosen 1997), given the ionization potential (Fuke et al. 1993) and typical bandgaps of SNPs. The 
lifetimes appearing in Equations (6) through (9) are: $\tau_{1}$, the radiative lifetime of the singly-excited state; $\tau_{2}$, the radiative lifetime of the doubly-excited state; $\tau_{A}$, the lifetime of the 
doubly-excited state against Auger ionization; $\tau_{i}$, the lifetime of the neutral state against 
photoionization by single photons; and $\tau$, the lifetime of the ionized state before recapture of a 
free electron, leading to a return to the neutral state.  We assume that the recaptured electron will appear in the conduction band of the SNP, 
with subsequent radiative transition to the valence band.

The solutions of Equations (6) through (11) are given by the following four expressions:
\begin{eqnarray}
P_{0} = [1 + \tau_{1} (W_{1} + \frac{1}{\tau_{i}}) + \frac{W_{2} \tau_{1} (W_{1} + \frac{1}{\tau_{i}})}
{(\frac{1}{\tau_{2}} + \frac{1}{\tau_{A}})} + \tau (\frac{W_{2} \tau_{1}(W_{1}+ \frac{1}{\tau_{i}})}
{\tau_{A}(\frac{1}{\tau_{2}} + \frac{1}{\tau_{A}})} + \frac{1}{\tau_{i}})]^{-1}
\end{eqnarray}
\begin{eqnarray}
P_{1} = \tau_{1}(W_{1} + \frac{1}{\tau_{i}}) \cdot P_{0}
\end{eqnarray}
\begin{eqnarray}
P_{2} = \frac{\tau_{1} W_{2}(W_{1} + \frac{1}{\tau_{i}})}{(\frac{1}{\tau_{2}} + \frac{1}{\tau_{A}})} \cdot P_{0}
\end{eqnarray}
\begin{eqnarray}
P_{+} = \tau[\frac{\tau_{1} W_{2}(W_{1} + \frac{1}{\tau_{i}})}{\tau_{A}(\frac{1}{\tau_{2}} +
\frac{1}{\tau_{A}})} + \frac{1}{\tau_{i}}] \cdot P_{0}
\end{eqnarray}

\noindent The ionization rate per SNP for single-photon ionization can be found from $P_0/\tau_{i}$, while the 2-photon Auger ionization rate becomes $P_{2}/\tau_{A}$. These two rates equal each other, provided

\begin{eqnarray}
\tau_{1} = \frac{(\tau_{A}(\frac{1}{\tau_{2}} + \frac{1}{\tau_{A}}))}{\tau_{i}W_{2}(W_{1}+\frac{1}{\tau_{i}})}.
\end{eqnarray} 

\subsection{Numerical Solutions}

For the numerical evaluations of these solutions, we need to find values for the rate coefficients and 
time constants involved in Equations (12) through (15).
We adopt the direction-integrated intensity of the interstellar radiation field
at the galactocentric distance of the Sun given by Mathis et al.(1983) and the optical absorption cross 
sections of SNPs by Kovalev et al. (2000a), extrapolated into the ultraviolet, to estimate the 
radiative excitation rate in the diffuse interstellar medium. We find 
$W_{1} \simeq 1.0 \cdot 10^{-5} \rm s^{-1}$ for nanoparticles with diameters near 3.5 nm under these conditions, and, as 
discussed above,
we set $W_{1} = W_{2}$. The timescale $\tau$ for recombinations between ionized SNPs and free electrons is estimated from the 
work of Draine \& Sutin (1987). With a characteristic diameter of 3.5 nm for a SNP with a single positive charge, an average electron 
density of $n_{e} = 0.03$ $\rm cm^{-3}$ from pulsar dispersion measures (Spitzer 1978), and an electron temperature $T_{e}$ = 
100 K we find $\tau \simeq 1.0 \cdot 10^{6}$ s, which includes an enhancement of the particle cross section due to
Coulomb focussing by a factor of about 200.
The timescale for single-photon 
ionization, $\tau_{i}$, cannot be estimated directly in the absence of measured ionization cross 
sections. However,
the observation of a large ERE efficiency in the diffuse ISM suggests a fairly
high ratio of neutral to ionized SNPs in this environment, and we estimate that 
$P_{0} \sim 0.5$. Equation (12) then leads to the conclusion that 
$\tau_{i} \simeq 1.0 \cdot 10^{6}$ s for
the diffuse ISM. Since single-photon ionization is the dominant photoionization process of SNPs in the 
diffuse ISM, we will use this result to estimate the cross section for this process later in Sect. 4.2.
Moreover, as the radiation density increases, the single-photon ionization rate
$\frac{1}{\tau_{i}}$ as well as the photoexcitation rates $W_{1}$ and $W_{2}$
scale in direct proportion of the radiation density. The radiative lifetime of the once-excited 
state, $\tau_{1}$, is strongly dependent upon size and temperature of the SNP and relatively long, consistent with the 
indirect-bandgap nature of silicon. From Takeoka et al. (2000) we estimate $\tau_{1}=2 \cdot 10^{-3}$ s for an 
appropriate average. The dynamics of the doubly-excited state is controlled by the small value of 
$\tau_{A} \sim 10^{-10}$ s (Efros \& Rosen 1997). We will assume that radiative decay from the doubly-excited 
state is not entirely prohibited by setting $\tau_{2} = 10^{-6}$ s.

We computed the probabilities $P_{0}, P_{1}, P_{2}$ and $P_{+}$ for a range of radiation densities extending from 
the interstellar density to $1 \cdot 10^{6}$ times the interstellar value. We also varied the lifetime of 
the ionized state, $\tau$, 
from the interstellar value to $1 \cdot 10^{6}$ times shorter, in steps of factors of ten. $\tau$ scales in 
inverse proportion to the electron density when the single-charge state is the 
dominant state among ionized SNPs, but $\tau$ can be much longer than expected,
if the bulk of SNPs are multiply ionized. At low electron energies, $\tau$ is relatively independent of the electron 
temperature (Draine \& Sutin 1987), as the temperature dependences of electron velocity and recombination cross section 
cancel one another. Aside from a constant factor, which measures the unknown fraction of the SNP absorption to 
the total absorption by all interstellar grains, the probability $P_{0}$ measures the intrinsic ERE quantum 
efficiency. Similarly, the probability $P_{1}$ is directly proportional to the ERE intensity. Finally, 
the ratios $\frac{P_{0}}{\tau_{i}}$ and $\frac{P_{2}}{\tau_{A}}$ measure the rates of single-photon ionization and two-photon Auger ionization, respectively.

In Figure 5 we plot the probability $P_{1}$ as a function of the radiation field density superimposed 
upon our ERE intensity data, compiled in Section 2. The model curves are normalized to fit the ERE 
intensity in the diffuse ISM at $log(U)$ = 0. The model provides a consistent explanation for all 
observed ERE intensities and upper limits by suggesting that the more intense radiation fields in 
reflection nebulae and HII regions will lead to higher ERE intensities, as long as the resultant 
increased photoionization rate is at least partially balanced by increased electron capture rates. The 
model fit suggests that in reflection nebulae the rate of returning ionized SNPs to the neutral state is up to one order of 
magnitude higher than in the diffuse ISM, consistent with the increased
densities and the absence of hydrogen ionization in these objects. In the Orion Nebula, electron capture rates higher
by at least five orders of magnitude than in the ISM are expected, consistent with electron temperatures of $T_{e} \sim 10^{4}$ K and electron 
densities near $10^{4}$ $\rm cm^{-3}$ generally quoted for the Orion Nebula (Rubin et al. 1998; Baldwin et al. 
1996). However, Figure 5 shows that the rate of returning ionized SNPs to the neutral state, $\frac{1}{\tau}$, increases by only three orders of magnitude
between the ISM and the Orion Nebula. We attribute this smaller than expected increase to
the dominance of multiple charge states among ionized SNPs in the Orion Nebula (Weingartner \& Draine 2001; Bakes \& Tielens 1994). Thus, most electron captures by ionized SNPs involve transitions between different charge states
instead of transitions between the singly-charged state and the neutral state.
It is of interest to note that only a fraction of about  1\% neutral SNPs is needed in order
to fit both the reflection nebulae and Orion Nebula intensity data.
Despite this low fraction of SNPs capable of luminescing, the ERE intensities are nevertheless high, 
because  (see Equation 3) the increase in the radiation field densities 
(factors of $10^{3}$ to $10^{5}$) more than balance the decrease in the efficiency (factor of $\sim 50$). In the Orion Nebula, a neutral SNP luminesces approximately once per second, while in the diffuse ISM only one 
luminescence event per neutral SNP occurs every 30 hours, on average.

\begin{figure}[th]
\begin{center}
\epsscale{0.7}
\plotone{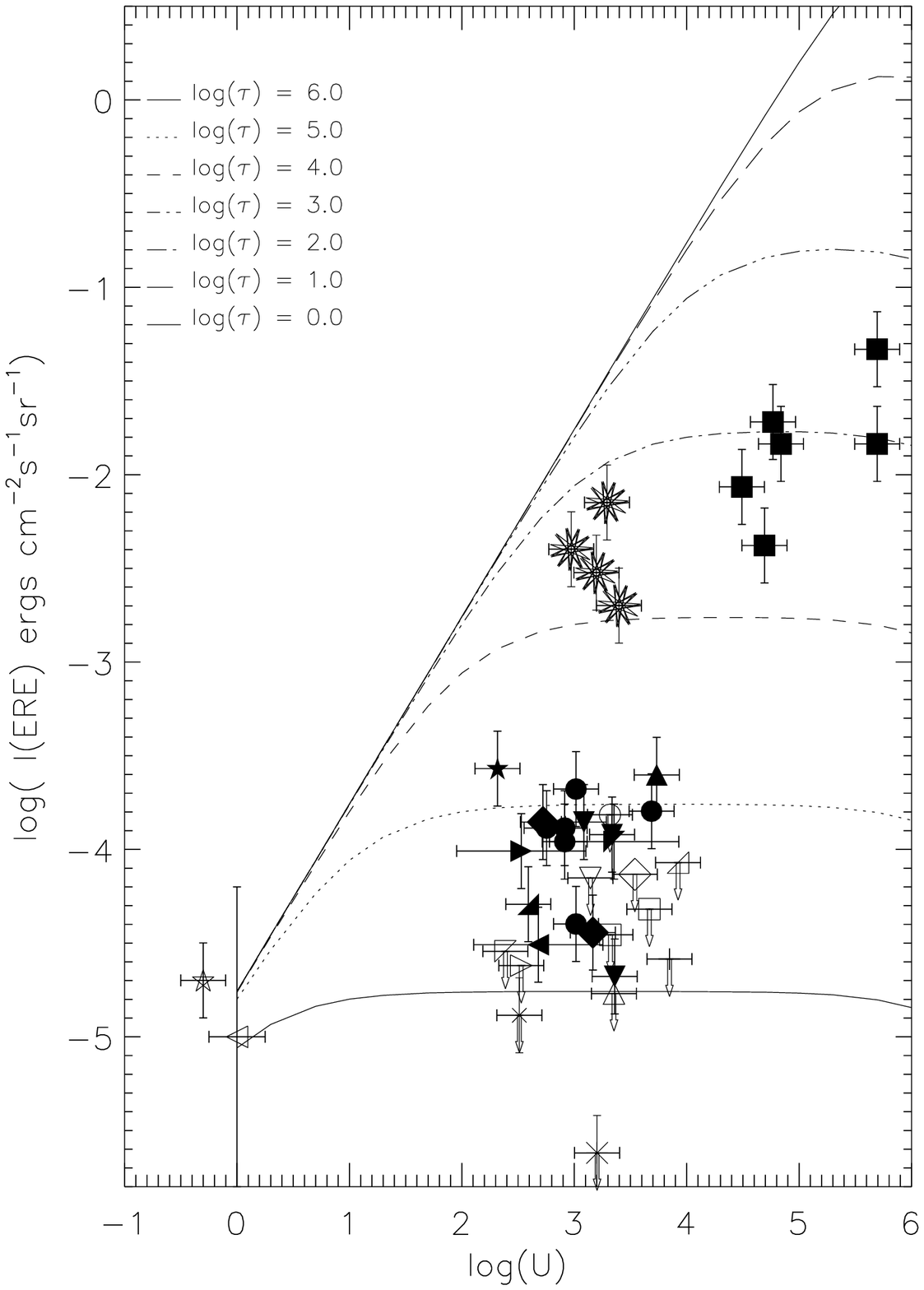}
\end{center}
\caption{The probability $P_{1}$ as a function of radiation field density, superimposed upon our 
ERE intensity data compiled in Section 2. The variable parameter $\tau$ measures the lifetime of a SNP in the ionized state in seconds.}
\end{figure}

In Figure 6 we plot the probability $P_{0}$ as a function of the radiation field
density superimposed upon our ERE efficiency data produced in Section 2.
Again, the model curves are normalized to fit the derived ERE efficiency for the diffuse ISM. The model 
matches the precipitous decline of the ERE quantum efficiency seen in reflection nebulae in the 
range $2 < log(U) < 4$
exceedingly well. With the illuminating stars in 
reflection nebulae having spectral types B 1 V or later, these nebulae, with H and He in the 
neutral state, form a sequence of low-electron-density environments. The Orion Nebula, by contrast, 
maintains a comparatively high ERE quantum efficiency, because its higher electron density 
effectively balances the ionizing effect of the more intense radiation upon the SNPs. The geometry of the Orion Nebula does not permit a reliable determination
of whether the observed ERE in this object is produced in the ionized region
or in the PDR on the face of the molecular cloud. However, the spatial distribution of the ERE detected in the HII region Sh 152 (Darbon et al. 2000)
suggests a close association of the ERE with the ionized volume.

\begin{figure}[th]
\begin{center}
\epsscale{0.9}
\plotone{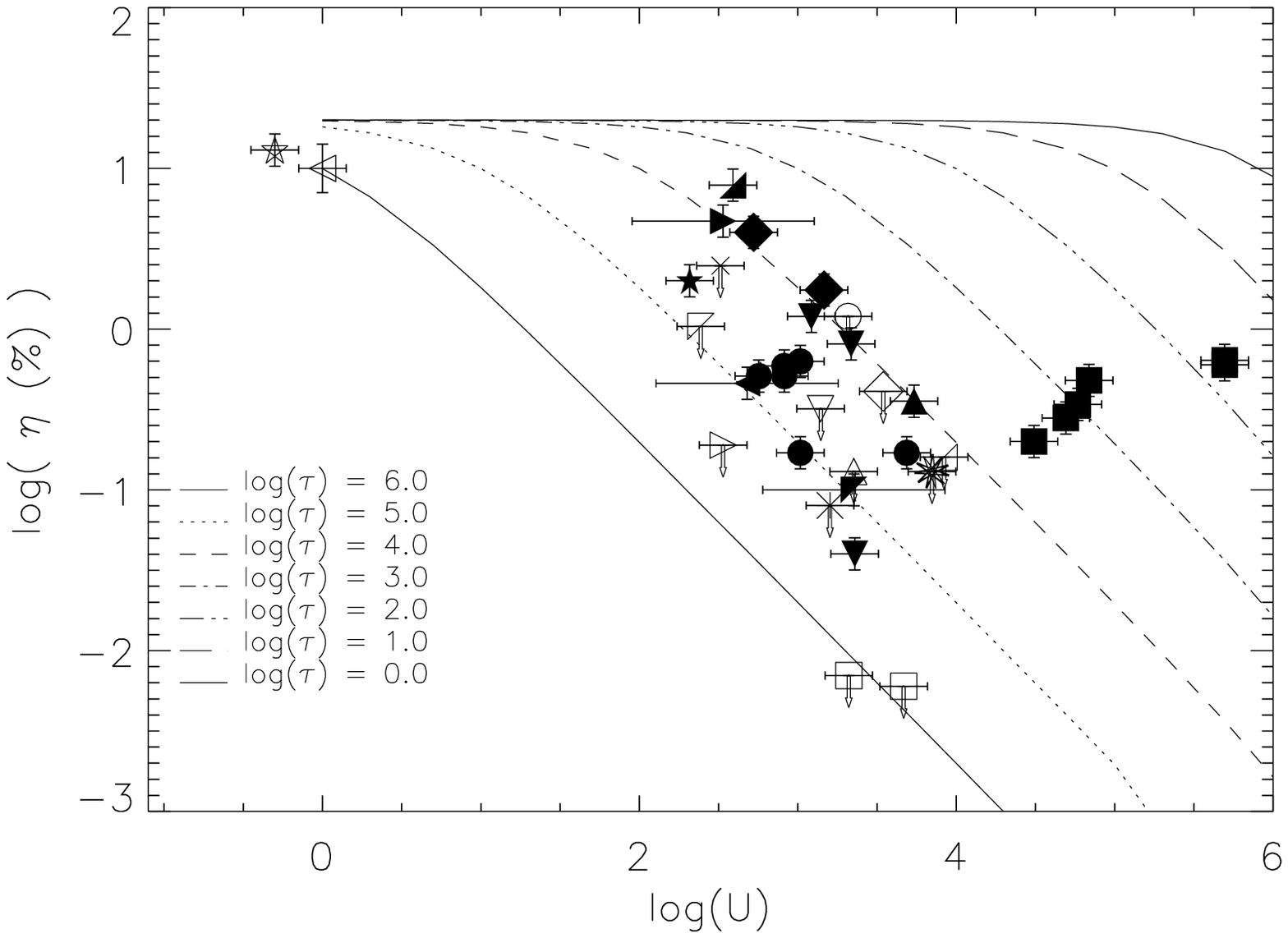}
\end{center}
\caption{The probability $P_0$ plotted as a function of the radiation
field density, superimposed upon the efficiency estimates of Section 2. }
\end{figure}

The comparison of the two SNP ionization processes is shown in Figure 7,
where the predicted ionization rates per SNP are plotted as a function of the radiation field density. 
Two sets of curves are shown for values of the parameter $\tau$ differing by five orders of 
magnitude.  Since the recombination of ionized SNPs with free electrons is independent of how 
a SNP was ionized, the ratio of the two ionization rates
is dependent only upon the radiation density and not on $\tau$.
We find that the single-photon ionization process clearly is dominant under
all conditions of radiation density found in the ERE sources studied here.
For the two-photon Auger process to become important under interstellar conditions, the lifetime of the excited state in SNPs, $\tau_{1}$, would need to be increased from the experimentally observed value by about 3 orders of magnitude, i.e $\tau_{1}$ $\simeq$ 2 s.

\begin{figure}[th]
\begin{center}
\epsscale{0.9}
\plotone{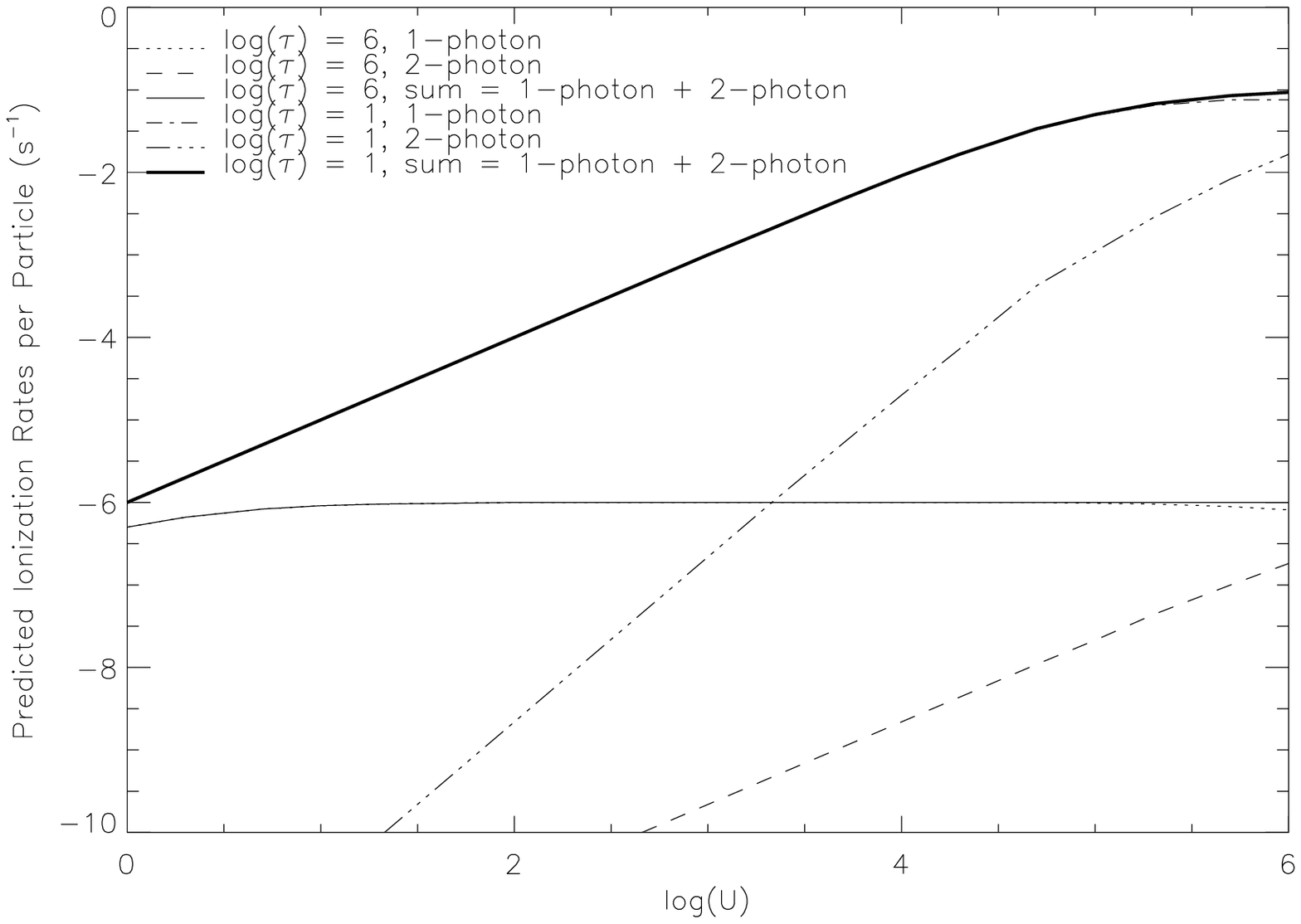}
\end{center}
\caption{The predicted ionization rates per SNP plotted as a function of the radiation
field density. Two extreme cases for low and high neutralization rates are shown.}
\end{figure}

\subsection{ The Size-Dependence of the Ionization Equilibrium}

The ERE band exhibits a considerable width, which is interpreted by the
the SNP model as reflecting the size distribution of the actively luminescing particles. The observations of the shift of the peak of the ERE intensity
toward longer wavelengths with increasing radiation density (Figure 2), when combined with the relationship between particle size and luminescence 
wavelength imposed by quantum confinement (Ledoux et al. 1998, 2000), imply
that this active size distribution is progressively eroded by radiation, 
starting with the smallest particles. While the single-photon ionization cross section has yet to be measured, we expect that it is large enough to make 
single-photon ionization of SNPs dominant over two-photon Auger ionizaton
under astrophysical conditions (Figure 7). We expect the size dependence of single-photon ionization of SNPs to involve a scaling with the geometric
cross section $\sim a^{2}$, despite the fact that for particles with radius 
$a \ll \lambda$, the photon absorption cross section $C_{abs} \sim a^{3}$ (Bohren \& Huffman 1983, p. 140). The reason for this difference lies in the fact that
electron escape lengths are small compared to the diameter range of the SNPs 
assumed responsible for the ERE (Weingartner \& Draine 2001). Ionization experiments with single
193 nm (6.4 eV) photons on neutral silicon nanocrystals (Ehbrecht \& Huisken 1999) have demonstrated the preferential ionization of the larger nanoparticles 
in a given, narrow, size distribution, covering the cluster size range from
1000 to 2000 atoms. According to Fuke et al. (1993), silicon clusters with more than 200 atoms have a first-ionization potential essentially equal to that of bulk silicon. Thus, the preferred ionization of larger clusters is not a
result of a lower ionization potential. Therefore, if single-photon ionization is the only process considered as controlling the size distribution of actively luminescent
SNPs, we conclude that the peak emission should shift progressively toward 
\emph{shorter} wavelengths with increasing radiation density as a result of of the preferential ionization of larger particles, the opposite of
what is being observed.

If the SNP model is to be maintained, other radiation-driven processes must
become important in high-radiation-density environments, which lead to the
efficient elimination of the smallest SNPs. We propose that photofragmentation
of multiply-charged SNPs, combined with single-photon heating, could be such a 
process. Our determinations of the ERE efficiency (Figure 4) suggest that the
fraction of neutral SNPs in RN is about 5\%, and about 1\% in the Orion Nebula.
Calculations of the expected charge equilibrium of 3 nm diameter particles
in these radiation environments predict distributions with peaks of multiple charge states of two to six positive charges per particle (Bakes \& Tielens 1994; Weingartner \& Draine 2001). Our own determinations of the lifetime of
the ionized state, single or multiple, in RN and in the Orion Nebula (Sect. 3.4)
also indicated a likely predominance of multiple positive charge states.

The balance between photoionization and electron capture by ionized nanoparticles in any given environment determines their charge distributions.
For a particle of given radius, the resulting charge distribution is almost entirely determined by the parameter U$\frac{\sqrt{T}}{n_{e}}$ (Bakes \& Tielens 1994; Weingartner \& Draine 2001). Unfortunately, for most of our ERE sources, there are no independent estimates of the temperature T or the electron
density $n_{e}$ of sufficient reliability to permit a valid examination of
the variations of $\lambda_{p}(ERE)$ and $\eta$ with U$\frac{\sqrt{T}}{n_{e}}$.
Of the two observed ERE parameters, $\eta$ represents additional complications, because its value is expected to depend also on the fractional abundance
of SNPs relative to larger grains, in addition to the expected dependence
on the fraction of SNPs remaining neutral. There is convincing evidence that the 
abundance of transiently heated particles, i.e. nanoparticles, varies substantially from cloud to cloud (Boulanger et al. 1990). This is in contrast to $\lambda_{p}(ERE)$, which only depends on the size distribution of the actively luminescent particles, not on their abundance. Therefore, in Figure 8
we employ generic environmental parameters for different ERE source environments to demonstrate the variation of $\lambda_{p}(ERE)$ with U$\frac{\sqrt{T}}{n_{e}}$
expected on the basis of our model. The control of the actively luminescent
size distribution by photon interactions is different in two separate domains.
In the first domain, defined physically by dense dark nebulae (DN) and the
diffuse ISM (DISM), increasing photoionization shifts the charge distribution
from a predominatly neutral state to a state roughly equally divided between
neutral particles and singly-charged particles. In this regime, the ERE peak wavelength is expected to shift towards shorter wavelengths as shown in Figure 8. In the second domain, consisting of reflection nebulae (RN) and HII-regions (HII), particles with two or more positive charges appear. Photofragmentation
of the smallest of such particles will rapidly shift the peak of the size distribution to larger sizes, resulting in larger values of $\lambda_{p}(ERE)$,
for the following reasons.

\begin{figure}[th]
\begin{center}
\epsscale{0.9}
\plotone{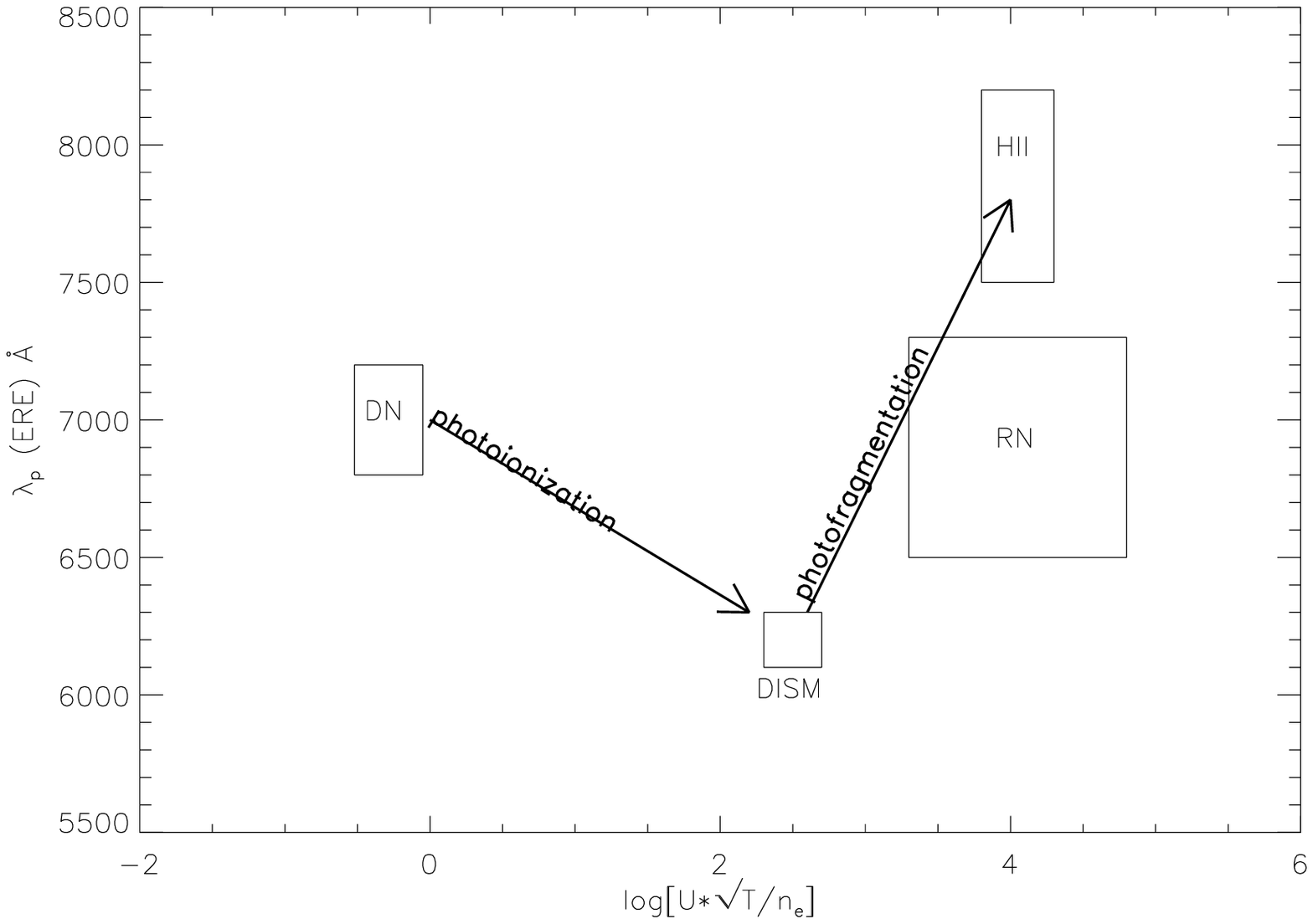}
\end{center}
\caption{The ERE peak wavelengths, as observed in high-latitude dark nebulae (DN), in the diffuse interstellar medium (DISM), reflection nebulae (RN), and HII-regions, are plotted against the parameter U$\frac{\sqrt{T}}{n_{e}}$, 
which largely determines the distribution of the carriers over positive charge states. The actively luminescent portion of the size distribution of ERE carriers is first shifted towards smaller sizes by photoionization affecting
predominantly the larger particles , then towards larger sizes by photofragmentation, which destroys smaller particles first.}
\end{figure}

The phase diagram of multiply-charged nanoparticles of a given positive charge
state recognizes three possible size ranges (Li et al. 1998). Below a certain critical size, multiply-charged nanoparticles will disintegrate via Coulomb
explosion even at 0 K and are thus unstable. Above this critical size, we encounter a range of diameters, where multiply-charged particles are metastable
and where fragmentation occurs as a result of Coulomb repulsion and heating to temperatures 0 K $<  T  <  T_{c}$, where $T_{c}$ is the critical
temperature, corresponding to the size-dependent boiling temperature of the 
nanoparticles (Li et al. 1998, eqn. 11). Beyond this metastable size range, particles of a given charge state are stable and can only be vaporized by raising their temperature to
$T > T_{c}$, the same as uncharged particles. Photofragmentation experiments
of multiply-charged silicon nanocrystals in the 1 to 6 nm size range 
(Ehbrecht \& Huisken 1999; Bescos et al. 2000) provide strong evidence that
doubly and triply ionized silicon particles in this size range are in the
metastable regime with respect to photofragmentation. Such particles decay by
ejecting small, singly-charged ionic fragments, predominantly $\rm Si^{+}$
monomers and $\rm Si_{2}^{+}$ dimers, thus achieving a state of higher stability.
However, this decay occurs only during the short interaction with a heating laser photon field and ceases immediately after the particles cool.

Under conditions present in astrophysical ERE sources, where photon fields are relatively diluted, the processes of excitation, ionization, and heating must be
considered as sequential, time-separated events, involving single photons.
In the diffuse ISM, the smallest stable SNP is a neutral particle which can survive the absorption of a 13.6 eV photon, emit a 2 eV ERE photon, and endure
a temperature pulse associated with the addition of 11.6 eV of internal energy.
This excess energy can also lead to an ionization, but each successive ionization
increases the ionization potential of the particle to (Seidl et al. 1991)

\begin{eqnarray}
I(z,a) = W_{b} + (2z - 1) \frac{e^{2}}{2a}  ,
\end{eqnarray}

\noindent where $W_{b}$ is the work function of the bulk material and where the second right-hand term represents the additional
energy required to remove a further electron from a (z-1)-fold positively charged, conducting sphere of radius $\rm a$. Photon 
absorptions by progressively multiply-charged SNPs, which are unable to
lose energy via luminescence and which are progressively more difficult to
ionize, lead to progressively increased single-photon heating. Recently,
Li \& Draine (2001a, 2001c) have calculated the expected temperature distributions of
charged pure silicon nanoparticles and of SNPs with $SiO_{2}$ mantles. They
predict maximum temperatures of $\sim 700$ K and of $\sim 380$ K, respectively, for particles of 2 nm diameter in a typical RN environment, such as NGC 2023, where multiple positive charge states would be expected for SNPs (Weingartner \& Draine 2001). Particles of twice this diameter reach
maximum temperatures of only about half these values under the same conditions.

In the light of these findings, we anticipate, therefore, that multiply charged,
metastable SNPs, upon heating by energetic UV photons, achieve a state of increased stabilty by ejecting singly-charged ionic fragments, thus reducing their overall charge state and mass. This effect will be strongest for the smallest SNPs, which, for a given charge state, experience the highest Coulomb
instability and which also gain the highest temperature increase upon absorption
of a photon of given energy. The gain in increased stability resulting from the ejection of an ionic fragment is only temporary, however, because a renewed ionization to the previous charge state will leave the now less-massive 
particle in an even less-stable state. Consequently, in dense radiation fields, 
the small-size end of an SNP size distribution will be progressively destroyed 
by a combination of Coulomb explosion and thermal heating. The only particles
permanently stable are the larger SNPs, which never reach charge/temperature 
combinations to place them into the metastable regime for photofragmentation.

\section{DISCUSSION}

\subsection{Limits on the Intrinsic Luminescence Efficiency of Interstellar SNPs}

The fit of our model calculations for the SNP quantum yield to the observed lower limits of the ERE luminescence 
efficiency was achieved by assuming that the intrinsic quantum yield of the SNP ensemble, under conditions present in the 
diffuse interstellar medium, is about 50\%. This number is obtained by having an even balance between neutral SNPs, 
which have an intrinsic quantum yield of 100\%, and ionized SNPs, which do not contribute to the ERE. Quantum 
yields between 50\% and 100\% have indeed been found for SNP laboratory samples (Wilson et al. 1993;
Credo et al. 1998; Ledoux et al. 2001). It is straightforward to show that the interstellar ERE carrier must have 
similar efficiencies.

Gordon et al.(1998) estimated that the ERE represents about 3\%
of the energy of the total dust emission in the high-Galactic-latitude ISM.
From a study of diffuse infrared Galactic emission with IRAS, Boulanger \& Perault (1988) concluded that about 40\% 
of the Galactic dust emission is attributable to dust grains small enough to undergo significant temperature 
fluctuations upon absorption of individual photons, which would include nano-sized ERE carriers. While cloud-to-cloud 
variations of this fraction were observed 
(Boulanger et al. 1990), this general result was confirmed by observations with DIRBE (Dwek et al. 1997; Bernard et 
al. 1994). A certain fraction, estimated as about 25\% of the small-particle emission or about 10\% of the total 
dust emission, is usually attributed to polycyclic aromatic hydrocarbons (PAHs) in order to account for the presence 
of the so-called unidentified infrared band (UIB) features in the Galactic spectrum (Desert et al. 1990; Dwek et al. 
1997; Misselt et al. 2001). If SNPs are the ERE carrier, they would contribute to the remaining 
small-particle emission component, which amounts to about 30\% of the total 
dust emission, after subtracting the PAH component. Other possible contributors
to the small-particle emission are tiny carbonaceous grains and tiny amorphous silicate grains (Li \& Draine 2001b).

Given the absorption spectrum of SNPs (Zubko et al. 1999) and the spectrum of the interstellar radiation field 
(Mathis et al. 1983), a typical photon exciting
an SNP has a likely energy of $E_{abs}$ = 8 eV, of which about 2 eV would be emitted as ERE, if the SNP is capable of 
luminescing, and 6 eV would be thermalized by the particle and emitted in the infrared. Let us consider two 
limiting cases for the intrinsic efficiency of the ERE carrier. The highest value for the quantum yield is 100\%, 
assuming that a single excitation can at most result in a single luminescence photon. Then, three times as much energy 
as is emitted in form of the ERE is contributed by thermal emission to the IR. With the estimate of Gordon et al.(1998), 
we find that 
about 9\% of the total dust emission, or about 1/3 of the small particle emission not attributed to PAHs could be 
resulting from the presence of SNPs in the interstellar dust mixture. Conversely, a lower limit to the intrinsic efficiency 
of the interstellar ERE carrier, assuming they are nanoparticles, can be obtained by assuming that all the 
small-particle thermal emission not assigned to PAHs is produced by the ERE carrier. The associated efficiency 
can then be obtained from (Duley 1992):

\begin{eqnarray}
\frac{ERE}{Thermal} & \sim & \left(\frac{3\%}{30\%}\right) \sim \left(\frac{\eta \cdot E_{ERE}}{E_{abs}
- \eta \cdot E_{ERE}}\right)
\end{eqnarray}

\noindent where $E_{ERE}$ is the bandgap energy of the ERE carrier. The resulting lower limit to the intrinsic efficiency of the ERE carrier is 
$\eta$ $\geq$ 36\%. Since our SNP model equates the intrinsic efficiency of the SNP ensemble with the fraction of 
neutral SNPs, our assumption of 50\% for this fraction is consistent with the observed characteristics of the 
actual ERE carrier.

\subsection{Prediction of the SNP Cross Section for Single-Photon Ionization}

An intrinsic quantum yield of $\geq 36\%$ of the ERE carrier in the diffuse ISM
implies that the value of $P_{0}$ (Eqn. 12) is $\geq$0.36 for the radiation field and the electron recombination rate 
applicable to that environment. This condition is met provided 
$\tau$ $\geq$ 1.8 $\tau_{i}$. Thus, we find:

\begin{eqnarray}
\frac{\tau}{\tau_{i}} & = & 
\left(\frac{ \langle \sigma \rangle _{i} \cdot \int_{5.1 eV}^{13.6 eV} 4 \pi J_{\nu} \frac{d\nu}{h\nu}} 
{ \pi a^{2} \cdot \tilde{\sigma} \cdot n_{e} \cdot v_{e} \cdot s}\right) \geq 1.8  ,
\end{eqnarray}

\noindent where $\langle \sigma \rangle_{i}$ is the average cross section for single-photon ionization, averaged over the energy range 
from 5.1 eV to 13.6 eV of the interstellar radiation field. Eqn.(19) allows us to evaluate the value of 
$\langle \sigma \rangle_{i}$.
With a geometric cross section of a typical SNP, $\pi a^{2} = 10^{-13}$ $\rm cm^{2}$, $\tilde{\sigma}$, the reduced cross 
section of Draine \& Sutin (1987) estimated at 200, an electron density $n_{e}$ = 0.03 $\rm cm^{-3}$, an 
average electron speed $v_{e} = 3 \cdot 10^{6}$ $\rm cm$  $\rm s^{-1}$, a sticking coefficient \emph{s} = 0.3,  and a flux of ionizing 
photons of 
$2.9 \cdot 10^{8}$ $\rm cm^{-2} s^{-1}$, we find 
$ \langle \sigma \rangle _{i}$  $\leq$  3.4 $\cdot 10^{-15}$ $\rm cm^{2}$. A published measurement of this quantity does not exist and our value is, 
therefore, 
a specific prediction made by the SNP model which can be used to either confirm or falsify our model. Our 
prediction applies to SNPs of average diameter 3.5 nm, expected to luminesce with peak emission around a 
wavelength of $\sim$ 700 nm (Ledoux et al. 2000). This ionization cross section is expected to vary with radius \emph{a} as 
$\sim a^{2}$ (see Sect. 3.5).

The photoionization cross section measures the likelihood that the absorption
of a single photon with energy above the ionization threshold of SNP's ($\sim$ 5.1 eV) leads to the ejection of an 
electron. Experiments conducted with palladium nanoparticles (Schleicher et al. 1993), with sizes and work 
function comparable to those of SNPs (Fuke et al. 1993), present useful information with which to assess the 
likelihood that our cross section prediction is reasonable. The 
absolute photoelectron yield per photon of palladium nanoparticles for photons within 1 eV above the threshold is only of the order of $10^{-2}$, and it 
approaches unity only at a photon energy of $\sim$ 10 eV. Similar increases in the photoelectric yield per photon by two to three 
orders of magnitude over the photon energy range from 5 eV to 10 eV have been reported by Burtscher et al. (1984) 
for small silver and gold particles. Theoretical calculations of the photoelectron yield of small silicate grains 
by Ballester et al. (1995) predict a yield of only 0.1 at the Lyman limit, decreasing to 0.001
at a photon energy of 8 eV. SNPs more likely will behave like silicates or other
insulators, where electron-phonon scattering interferes with the ejection of photoelectrons, rather than like 
small metal clusters, where electron-electron scattering takes the place of the process by which most of the 
energy of absorbed photons with energy $E > E_{i}$ is converted into heat (Gail \& Sedlmayr 1980). This suggests, 
if the cited results are applicable to SNPs, that the photoionization cross section of SNPs is substantially 
smaller than the absorption cross section throughout most of the astrophysically relevant UV spectrum. Absorption 
cross sections measured for SNPs (Kovalev et al. 2000) cover only the photon energy range from 1.48 eV to 3.53 eV 
and extend from a few $10^{-19}$ $\rm cm^{2}$ to about $10^{-14}$ $\rm cm^{2}$, enough to estimate that the absorption cross 
sections of 
SNPs of about 3.5 nm diameter are not more than about $10^{-13}$ $\rm cm^{2}$ in the UV. According to our estimate, the 
single-photon photoionization cross section of SNPs, then, is not greater than about $3 \cdot 10^{-2}$ times the absorption cross section.

Another way to assess the likely validity of our result is to extrapolate the ionization cross section of SNPs from the ionization cross section of neutral 
silicon atoms. Near threshold, the latter cross section is about 5 $\cdot 10^{-17}$ $\rm cm^{2}$ (Verner et al. 1996).
A typical SNP of 3.5 nm diameter has about 800 silicon atoms, with a combined ionization cross section of 4.0 $\cdot 10^{-14}$ 
$\rm cm^{2}$. Our estimate is about one order of magnitude smaller, corresponding to a photoelectron yield of about 0.1. This 
agrees surprisingly well with the yields found in experiments and in calculations, when averaged over the energy range of 
photons from 5.1 to 13.6 eV and a spectrum given by that of the interstellar radiation field.

\subsection{The Special Role of Lynds 1780}

Lynds 1780 (L 1780) is a high-latitude (b = $36.9^{\circ}$) dark nebula. Mattila (1979) noted the unusually red color of 
this object and determined its spectral energy distribution, but given that ERE as a general interstellar phenomenon 
was not to be recognized for several more years, he failed to find a likely source for the apparent red excess in 
the spectrum of L 1780. The identification of the red excess with ERE was proposed by Chlewicki \& Laureijs (1987), 
who noted the similarities between the spectra of L 1780 and the Red Rectangle. 
L 1780 is an optically thick cloud 
($A_{B} \sim$ 3 mag) with a density of about $10^{3}$ $\rm cm^{-3}$ (Mattila 1979), exposed to the  local (D $\sim$ 100 pc) 
interstellar radiation field.

L 1780 occupies a special role among all ERE sources with likely interstellar dust mixtures: it has the highest 
efficiency determined so far ($\sim$ 13\%; Fig. 4) for an object containing  such a dust mixture and it exhibits a very broad emission band with a peak near 700 nm 
(Fig. 2).
Until now, in view of the low density of the exciting radiation field this relatively long peak wavelength  made 
L 1780 appear the odd outlier on graphs such as Figure 2. Our SNP model provides a natural explanation for the 
seemingly odd characteristics of the ERE in L 1780. In view of the much higher gas density of L 1780 compared to the 
diffuse interstellar medium in general, the electron density will be higher in L 1780 as well, while the UV-photon density is actually reduced due to the cloud's intrinsic opacity, so that the overall 
ionization equilibrium tends more towards neutrality. Hence, $\frac{\tau}{\tau_{i}} \ll$ 1 in L 1780, 
while $\frac{\tau}{\tau_{i}} \sim$ 1 
in the diffuse ISM and $\frac{\tau}{\tau_{i}} \gg$ 1 in reflection nebulae and HII regions. These are the conditions, respectively, under which our model
predicts predominantly neutral SNPs, SNPs approximately equally balanced between neutral and
ionized, and predominantly ionized SNPs. Thus, in L 1780 the SNPs are more predominantly neutral than is the case in the diffuse ISM. The larger SNPs, which are removed from 
luminescing in the diffuse ISM due to the size-dependent ionization, are now restored as ERE contributors. Consequently, both the overall quantum 
efficiency increases and the peak wavelength shifts towards larger values compared to the diffuse ISM. The ERE 
spectrum of L 1780 may, therefore, be regarded as most closely approximating that
of a size distribution of luminescing SNPs, hardly modified by ionization.

\subsection{Multiple Ionization and Photofragmentation}

In Section 3.5 we have argued that Coulomb explosions of multiply-charged,
metastable SNPs will gradually erode the small-size end of an existing
SNP size distribution, once it is exposed to dense and hard radiation fields.
An observable consequence of such conditions would be the increased gas-phase
abundance of $\rm Si^{+}$ ions as well as SiO molecules in the photon-dominated
regions between dense molecular clouds and nearby hot stars.
An impressive example of such an enhanced 
gas-phase abundance of $\rm Si^{+}$ in the presence of a dense radiation field can be found in the ISO observations of the 
reflection nebula NGC 7023 by Fuente et al. (2000). In the inner cavity of this nebula, where ERE is almost totally 
absent (Witt \& Boroson 1990), at least 20\% to 30\% of the cosmic abundance of silicon is found in the gas phase, 
while the inferred gas-phase abundance of silicon drops to about 5\% in the high-density ERE filaments to the south 
and to the north-west of the central B 3 V star, HD 200775. This is
consistent with an almost total photodisintegration of all SNPs in this volume, given that the SNP model requires about 
20\% of the total silicon abundance in
the form of SNPs. We suggest that multiple ionization of SNPs and the 
resulting gradual photofragmentation may have destroyed the SNPs in the inner cavity of NGC 7023. We also suggest that this is the process by which the size
distribution of SNPs in regions of high radiation density is changed permanently, so that only the larger SNPs capable of luminescing at longer wavelengths survive.

\subsection{Infrared Band Emissions from SNPs}

A frequently raised question about interstellar SNPs concerns expectations that SNPs should give rise to infrared 
band emissions for which there is no clear
present evidence. Li \& Draine (2001a, 2001c) made a first attempt at calculating the expected emission spectrum of pure 
silicon particles and SNPs with oxide coatings. They predict the presence of a relatively sharp emission feature at 
16.4 $\mu$m for pure silicon particles and a still stronger, broad emission peak at 20 $\mu$m due to $\rm SiO_{2}$ for 
oxide-coated 
SNPs. Several 
comments are in order regarding pure silicon particles: i) The case of pure silicon particles appears astrophysically less interesting; ii) such particles, with dangling bonds at the surface remaining unpassivated, would not luminesce in the ERE band; iii) they have never been proposed as the ERE carrier; iv) they would 
not be expected to be chemically stable under
interstellar conditions, if they could form in the first place. In our view, the predicted
16.4 $\mu$m feature is, therefore, of questionable significance.

Adding an oxide coating to SNPs, according to Li \& Draine (2001a), lowers the 
most likely temperature of the nanoparticles from 300 K to about 75 K, due to the oxide's ability to radiate strongly 
in Si-O vibrational bands. This lower temperature, however, prohibits significant emission at the most commonly 
investigated band near 10 $\mu$m and leaves the radiation near 20 $\mu$m as the dominant mode. The detectability of 
the 20 $\mu$m
 band is a function of the SNP abundance as a fraction of the radiating dust mass, and thus is inversely proportional 
to the intrinsic ERE quantum efficiency of the oxide-covered SNPs. Experimentally determined quantum efficiencies for SNPs are found to be in the range from 50\% to 100\% (Wilson et al. 1993; Credo et al. 1999; Ledoux et al. 
2001). Li \& Draine (2001a), from their model calculations, conclude that
that even for such high-efficiency SNPs, the predicted intensity of the 20 $\mu$m band emission from the diffuse ISM exceeds observed upper limits set by
DIRBE observations by a factor of 7. Similarly, in the reflection nebula
NGC 2023, they find the predicted 20 $\mu$m feature strength to exceed the
observed intensity seen in an ISO spectrum by a significant factor. In order
to decide, whether these constraints are in fact as severe as they appear,
several objectives would need to be pursued in the future. These include
sensitive spectroscopic observations of the 20 $\mu$m region in ERE sources,
experimental determinations of the dielectric functions of oxygen-passivated
SNPs for more solidly-based model calculations, as well as controlled oxidization experiments aimed at determining the minimum amount of $\rm SiO_{2}$
needed for the effective passivation of SNPs.

Detection of the predicted 20 $\mu$m emission band would be most likely in objects where the abundance of SNPs exceeds 
that in the interstellar medium by a wide margin. The Red Rectangle (Schmidt et al. 1980) has an unusually strong 
ERE band, which suggests that the dust locally produced in this proto-planetary nebula is exceptionally rich in the 
particles producing the ERE. The publically accessible ISO-SWS spectrum of the Red Rectangle 
$\it{(http://isowww.estec.esa.nl/galleries/cir/red\_rect.html)}$ does not exhibit any prominent feature at 20 $\mu$m. 
However, a 
detailed analysis of this spectrum has yet to be done.
The Red Rectangle is a member of a class of similar objects, bi-polar proto-planetary nebulae with dust 
produced by mass-loosing AGB stars. Hrivnak et al. (2000) have used ISO to study the infrared spectra of nine such 
objects. Four of these exhibit a strong emission band at 21 $\mu$m, closely resembling the profile predicted by Li \& 
Draine (2001a). In two additional objects, the identification of the 21 $\mu$m band is tentative; in three objects the 
band was not detectable. None of these nebulae has been studied for the presence of ERE, and none is expected, 
because the spectral types of the
central stars are much later than that of the illuminating star in the Red Rectangle (B 9.5 III). The possibility 
that SNPs are formed in protoplanetary nebulae and are excited to emit ERE in the case where central stars with 
sufficient ultraviolet photons are present, deserves further investigation.
There has also been an indication that the 21 $\mu$m band is present in the dust emission spectrum of the supernova remnant 
Cas A (Douvion et al. 2001), which, together with the strong presence of SiO emission from the remnant of SN 1987A
(Roche et al. 1991), leads to the suggestion that supernova remnants could be locations for SNP formation. An alternate 
suggestion for the identification of the 21 $\mu$m feature in proto-planetary nebulae has been provided by von Helden 
et al. (2000), who proposed titanium carbide nanocrystals as the carrier of this feature. No estimates exist on whether 
the likely abundance of titanium carbide
is sufficient to explain the intensity of the 21 $\mu$m band.

\section{SUMMARY}

The existing body of observational data on the interstellar luminescence process known as ERE shows a wide range of 
variation in three observational characteristics: the band-integrated intensity, the wavelength of peak emission, 
and the efficiency with which absorbed UV/optical photons are converted into ERE photons. By systematically analysing 
all suitable ERE observations available so far, we have shown that the common environmental factor driving the variations in 
these three characteristics is the radiation field density in the different environments, where ERE has been 
detected. The ERE intensity increases with radiation density roughly linearly, as expected for photoluminescence; 
the peak of the ERE band shifts from about 610 nm in the diffuse ISM to beyond 800 nm
in O-star dominated HII regions; and the photon conversion efficiency is found to be highest in environments with the 
lowest density of exciting photons, only to decline significantly in denser radiation fields.
In the second part of this paper, we have introduced a model for the ERE carrier
in which photoionization of nanoparticles, balanced by recombination with free
electrons, is the controlling physical process which determines the total fraction of actively luminescing particles. The model relies on experimental findings, showing 
that ionization will quench photoluminescence in a luminescing semiconductor nanoparticle, with luminescence being restored upon 
recombination. This model successfully reproduces the observed variations of the
ERE intensity and the ERE photon conversion efficiency for a large range
of radiation field densities.
We find that under the complete range of conditions where ERE is being observed in astronomical environments, the single-photon ionization process is always  dominant over the two-photon Auger ionization process. In dense-radiation environments, photoionization will lead to 
relatively high positive charge states for SNPs, which will render the
smallest SNPs unstable against photofragmentation upon single-photon heating.
We suggest this process as the explanation for the observed shift of the
wavelength of peak ERE intensity toward larger values in  UV-radiation fields of
increased density.
Silicon nanoparticles have been proposed as carriers of the ERE in view of their ability to match the spectral 
characteristics of the ERE and the efficiency constraints better than any other current model candidate.  In order to match the entire body of ERE data presented in this paper, we require that 
silicon nanoparticles with an average diameter of 3.5 nm have an ionization cross section of about
$3.4 \cdot 10^{-15}$ $\rm cm^{2}$ when 
exposed to a radiation field with an energy distribution equal to that found in the diffuse ISM of the solar neighborhood.
The experimental determination of this cross section would represent a critical test for the model proposed here.
Finally, we note that sensitive spectroscopic searches for the 20 $\mu$m band
predicted to be associated with oxygen-passivated SNPs in ERE-bright interstellar environments would provide severe constraints for the validity
of the SNP model, especially if accompanied by experimental efforts to determine the dielectric functions for these particles in the laboratory.

Acknowledgement: We are extremely grateful to Bruce Draine, Friedrich Huisken,
Gilles Ledoux, and Aigen Li, who reviewed an earlier version of this paper and
who provided much critical input to the present version. We also acknowledge
extensive stimulating discussions about ERE and nanoparticles
with Louis Brus, Bob Deck, Walt Duley, Minoru Fujii, Dieter Gerlich, Karl Gordon, Thomas
Henning, and Daniele Pierini. Finally, we acknowledge again Bruce Draine, who as  referee provided further helpful criticism and several very constructive
suggestions. This work was supported through Grant NAG5-9202
from the National Aeronautics and Space Administration to The University of 
Toledo.

\end{document}